\documentclass[aps,prd,10pt,amsmath,amssymb,nofootinbib,notitlepage,superscriptaddress,twocolumn]{revtex4-2}

\usepackage[utf8]{inputenc}
\usepackage{amsmath,amssymb}
\usepackage{comment}
\usepackage{amsmath}
\usepackage{braket}
\usepackage{mathrsfs}
\usepackage{mathtools}
\usepackage{array}
\usepackage{fancyhdr}
\usepackage{xcolor}
\usepackage{bm}
\usepackage{url}
\usepackage{multirow}
\usepackage{makecell}

\usepackage{ragged2e}

\usepackage{diagbox}
\usepackage{nicematrix}
\usepackage{tikz}
\usetikzlibrary{patterns}
\usetikzlibrary{arrows.meta}

\allowdisplaybreaks

\definecolor{RedWine}{rgb}{0.743,0,0}
\definecolor{RoyalBlue}{rgb}{0.25,.41,.88}
\definecolor{ForestGreen}{rgb}{.13,.54,.13}

\newcommand{\aap}{A\&A}
\newcommand{\apjl}{Astrophys. J. Lett.}

\usepackage{tensor}
\usepackage{slashed}
\usepackage{dsfont} 
\usepackage{bbold}

\usepackage[backref,breaklinks,colorlinks,citecolor=ForestGreen]{hyperref}

\begin{document}

\title{Schwinger effect with backreaction in 1+1D massive QED with a strong external field}

\author{Samuel E. Gralla}
\affiliation{
Department of Physics, University of Arizona, Tucson, Arizona 85721, USA
}

\author{Morifumi Mizuno}
\affiliation{
Department of Physics, University of Arizona, Tucson, Arizona 85721, USA
}

\begin{abstract}
In the presence of a strong electric field, the vacuum is unstable to the production of pairs of charged particles---the Schwinger effect. The created pairs extract energy from the electric field, resulting in nontrivial backreaction.  In this paper, we study 1+1D massive QED subject to strong external electric fields in a self-consistent and fully quantum manner.  We use the bosonized version of the theory, which attains a cosine interaction term in the presence of nonzero fermion mass $m$.  However, the assumption of strong electric field justifies a perturbative treatment of the cosine interaction, i.e., an expansion in  $m$.  We calculate the vacuum expectation value of the electric field to first order in $m$ and show that---surprisingly---it satisfies a classical nonlinear partial differential equation (related to the sine-Gordon equation).  We show that the electric field exhibits dissipation-free oscillations (analogous to ordinary plasma oscillations) and calculate the plasma frequency analytically.  We also compare to the semiclassical approximation commonly used to study backreaction, showing that it fails to capture the $O(m)$ shift in the plasma frequency.
\end{abstract}
\maketitle

\section{Introduction}
The mere existence of sufficiently strong electric fields leads to the production of electron-positron pairs due to the Schwinger effect~\cite{Heisenberg:1936nmg,Schwinger1951jun}. 
The critical field strength for pair production is given in terms of the electron mass $m$ and elementary charge $e>0$ by 
\begin{align}
    \label{eq:Schwinger critical}
    E_{c}=\frac{m^{2}c^{3}}{e\hbar}\sim 10^{18}\mathrm{V/m},
\end{align}
below which the effect is exponentially suppressed. 
Particle creation leads to a natural question concerning the energetics of the mechanism: Where does the energy of those particles come from? 
Clearly, the electric field itself is the source; thus, the created particles must extract energy from the field and reduce its strength. In addition, the newly born charged particles will both feel and modify the electric field.  Understanding these complex dynamics is the problem of Schwinger effect \textit{backreaction}. Backreaction to particle creation is also important in gravitational problems, such as particle production by the expanding universe \cite{Parker:1968mv} and Hawking radiation (e.g., \cite{Birrell:1982ix,Parker:2009uva}).

One may study the backreaction to the Schwinger effect using the ``semiclassical" approximation, in which the electric field is treated as a classical field rather than as an operator.  Then, the classical field is demanded to be sourced by the expectation value of the current \cite{Cooper.Mottola1989jul,Kluger.Eisenberg.ea1992jun,Kluger.Eisenberg.ea1993jun,Kluger.Mottola.ea1998nov,Pla:2020tpq,Newsome:2024spi}.  The semiclassical approximation is formally justified in a large-$N$ expansion, but may still provide insight in the context of a single matter field.

Ultimately, the time evolution needs to be derived from full QED.  While the general case remains intractable, Chu and Vachaspati \cite{Chu.Vachaspati2010apr} showed that backreaction can be treated exactly for a massless fermion in 1+1D (the ``Schwinger model'' \cite{Schwinger1962dec}) using the method of bosonization \cite{Coleman1975apra,Mandelstam1975maya}.  
They used the fact that 1+1D massless QED is ``dual" to a free scalar (boson) theory of mass $M=q/\sqrt{\pi}$ ($q$ is the elementary charge in 1+1D) \cite{Lowenstein.Swieca1971nov,Manton1985jan,Iso.Murayama1990jul}, and showed that, by virtue of the linearity of the field equation, a fully quantum description of the screening process can be obtained by \textit{simply solving a classical Klein-Gordon equation}.

In the massive Schwinger model (1+1 QED with fermions of mass $m$), the bosonized theory is no longer free.  The interacting theory requires an explicit regulator to define, and the associated scale $\Lambda$ actually appears in the boson field equations.  If we include an external classical field $E_C$, then the field equations are 
\begin{align}
    \label{eq:boson equ??}
    &\left(\partial_{t}^{2}-\partial_{x}^{2}+\frac{q^{2}}{\pi}\right)\phi+\Lambda m \sin\left(\sqrt{4\pi}\phi\right)=-\frac{q}{\sqrt{\pi}}E_{C},
\end{align}
with the precise meaning of $\sin (\sqrt{4\pi} \phi)$ depending on how the theory is regulated.  The value of $\Lambda$ should not affect observable predictions, but any attempt to treat \eqref{eq:boson equ??} classically must somehow pick a value for $\Lambda$.  Prior work has fixed $\Lambda$ in a derivative expansion \cite{Jentsch.Daviet.ea2022jan,Batini.Kuhn.ea2024aug}, at large occupation number \cite{Gold.McGady.ea2021oct}, or in a WKB approximation \cite{Pham.Zager.ea2025apr}, or simply left it as a free parameter \cite{Blake:2012tp,Gralla2019maya}.

In this paper we will study the massive theory as a perturbation of the massless theory and show that a  version of \eqref{eq:boson equ??} indeed arises to first order in $m$.  Our field operators are defined in the Schr\"odinger picture, 
and regularization is done with normal ordering  \cite{Coleman1975apra}.  We assume that at early times the quantum state is the vacuum and the external classical electric field is zero. Then, the external field turns on at some time $t=t_{0}$ after which it varies arbitrarily.  
Under these assumptions, we find that, to first order in $m$, the field expectation value $\braket{\phi}$ satisfies the following equation:
\begin{align}
    \label{eq:PDE for <phi>}
    &\left(\partial_{t}^{2}-\partial_{x}^{2}+\frac{q^{2}}{\pi}\right)\braket{\phi}+\frac{e^{\gamma}mq}{\pi}\sin\left(\sqrt{4\pi}\braket{\phi}\right)=-\frac{q}{\sqrt{\pi}}E_{C},
\end{align}
where $\gamma=0.57721\cdots$ is the Euler–Mascheroni constant.
This is the sine-Gordon equation \eqref{eq:boson equ??} with the particular choice $\Lambda=e^{\gamma}q/\pi$, and $\phi$ is now understood as describing the expectation value of the field. We find it highly nontrivial that this classical-looking result arises from a fully quantum treatment. We emphasize in particular that no classical $\hbar \to0$ limit has been taken. Instead, Eq.~(\ref{eq:PDE for <phi>}) describes a truly quantum phenomenon.  

The boson expectation value $\braket{\phi}$ is related to the electric field expectation value by 
\begin{align}\label{eq:<E>}
    \braket{E}=E_{C}+\frac{q}{\sqrt{\pi}}\braket{\phi}.
\end{align}
Thus we see that the expectation value of the electric field obeys a simple partial differential equation (\ref{eq:PDE for <phi>}).  Notice that no ``particles" or ``pair creation" are explicitly visible in Eq.~(\ref{eq:PDE for <phi>}), and these concepts likewise do not arise in its derivation.  
We find this advantageous since the fundamental objects in QFT are the fields rather than particles. Eq.~(\ref{eq:PDE for <phi>}) simply shows the response of local observables (expectation value of $E$) to the external field. Its solutions can be used to validate the heuristic idea of created particles screening electric fields.

Indeed, this process was beautifully illustrated in the massless case by Chu and Vachaspati \cite{Chu.Vachaspati2010apr} in the setting of capacitor breakdown.   We revisit this problem in the (perturbatively) massive case and consider a slightly modified setup. We ensure local charge conservation for the external source by ``assembling'' the capacitor: its plates begin at the same location and then move to their final separation.  For the boundary condition of the field, we consider the two cases: 
(i) the field may freely pass through the plates, and 
(ii) the field outside the capacitor is strictly zero (as if the plates are mirrors).  For the first case, we found agreement with Refs.~\cite{Chu.Vachaspati2010apr,Gold.McGady.ea2021oct,Pham.Zager.ea2025apr}  that the electric field is screened inside the capacitor.  However, for the second case where the dynamical part of the field cannot spread out to infinity, the field never reaches a static state, but rather continues to oscillate indefinitely.

The frequency of these oscillations is the effective plasma frequency for the QED plasma created by the strong external field.  We may define a precise plasma frequency by considering spatially homogeneous external fields.  By solving \eqref{eq:PDE for <phi>} perturbatively, we find that the QED plasma frequency is given to $O(m)$ by
\begin{align}\label{eq: plasma frequency}
    \omega = \frac{q}{\sqrt{\pi}}+\frac{mqe^{\gamma}}{\pi 
    E_{C}}\cos{\left(\frac{2\pi E_{C}}{q}\right)}J_{1}\left(\frac{2\pi E_{C}}{q}\right).
\end{align}

We also study the problem semiclassically in order to test the validity of the semiclassical approximation.  Semiclassical results are available for spatially homogeneous electric fields (at any value of $m$) \cite{Ferreiro.Navarro-Salas2018jun} and can be compared to our small-$m$ results in full QED.  We find that the semiclassical approximation is exact at $m=0$, but fails at $O(m)$, predicting that there is no correction at this order, in disagreement with our full-QED result.  We may say that the semiclassical approximation is qualitatively correct but quantitatively incorrect, in particular failing to capture the $O(m)$ shift in the plasma frequency \eqref{eq: plasma frequency}.

As explained in Sec.~\ref{sec:expectation value}, our perturbative treatment of the fermion mass is justified at strong electric field $(E_C \gg m)$ and/or strong fermion coupling $(q \gg m)$.  However, versions of Eq.~\eqref{eq:boson equ??} with the same value $\Lambda=e^{\gamma}q/\pi$ also arise in other physical regimes \cite{Gold.McGady.ea2021oct, Batini.Kuhn.ea2024aug,Pham.Zager.ea2025apr}, suggesting a more fundamental significance within the massive Schwinger model.  Further research is required to clarify the scope of the quasi-classical description in massive 1+1D QED.

This paper is organized as follows. In Section \ref{sec:full QED}, we review the bosonization of the massive Schwinger model and derive the QED Hamiltonian in terms of the boson field.  In Section \ref{sec:expectation value}, we calculate $\widehat{\phi}$ to first order in the fermion mass and show that it satisfies Eq.~(\ref{eq:PDE for <phi>}).  
In Section \ref{sec: capacitor breakdown}, we revisit the capacitor breakdown problem and demonstrate that the screening effect is captured by solving Eq.~(\ref{eq:PDE for <phi>}).  In Section~\ref{sec:plasma frequency}, we consider a spatially homogeneous background field and derive the plasma frequency (\ref{eq: plasma frequency}).  In Section \ref{sec:QED and semiclassical}, consider the same problem in the semiclassical approximation and compare the results.  We conclude in Section \ref{sec:outlook} with a brief discussion of interesting future directions.

In this paper, we take $c=\hbar=1$, and the signature of the metric is $(-,+)$. 

\section{Bosonization of 1+1D QED}\label{sec:full QED}
In this section, we derive the QED Hamiltonian in terms of the boson field, closely following Coleman \cite{Coleman1976sep}.\footnote{In \cite{Coleman1976sep}, the signature of the metric is opposite to that in our paper.}
The Lagrangian of 1+1D QED in the presence of an external classical source $J^{\mu}_{C}$ satisfying the conservation law $\partial_{\mu}J^{\mu}_{C}=0$ is given by 
\begin{align}
    \label{eq:Lagrangian QED}
    L&=\int \! dx\left[
    i\overline{\psi}\gamma^{\mu}\partial_{\mu}\psi-m\overline{\psi}\psi+qA_{\mu}\overline{\psi}\gamma^{\mu}\psi
    \right.\notag
    \\
    &\hspace{3cm}\left.
    -\frac{1}{4}F^{\mu\nu}F_{\mu\nu}+A_{\mu}J^{\mu}_{C}
    \right].
\end{align}
\smallskip
The spatial integral is taken from $-\infty$ to $\infty$, and $\gamma^{\mu}$ are the gamma matrices, for which we take the following representation:
\begin{align}
    \label{eq:gamma matrix}
    \gamma^{t}=\begin{pmatrix}
        0&1\\1 &0
    \end{pmatrix},
    \quad 
    \gamma^{x}=\begin{pmatrix}
        0&1\\-1 &0
    \end{pmatrix}.
\end{align}
In 1+1D, the electromagnetic field behaves quite differently compared to the 3+1D case. 
First of all, there is no magnetic field, and there is only one component of the electric field. 
Furthermore, there are no electromagnetic waves since there are no transverse fields. 
This can be understood in a physical sense by considering a uniformly charged plate in 3+1D. The electric field created by the plate is uniform everywhere, and it never decays, no matter how far one goes away from the plate. In addition, the motion of the plate does not affect the electric field at all, i.e., no waves are generated. 
In the quantum theory, this indicates the non-existence of photons. 
This feature of the electromagnetic field in 1+1D can be most simply captured by taking the Coulomb gauge $A^{\mu}(t,x)=(A(t,x),0)$. We will use this gauge exclusively, and express our final result in terms of the gauge-invariant electric field. 
The electric field $E$ is then given by 
\begin{align}
    E(t,x)=-\partial_{x}A(t,x).
\end{align}
Then, the Lagrangian (\ref{eq:Lagrangian QED}) no longer depends on the time derivative of $A(t,x)$. 
Variation of Eq.~(\ref{eq:Lagrangian QED}) with respect to $A(t,x)$ results in  Gauss' law:
\begin{align}\label{eq:Gauss law}
    \partial^{2}_{x}A(t,x)=-(q\psi^\dagger\psi+J^{t}_{C}).
\end{align}
We express the general solution as
\begin{align}\label{eq:classcial A}
    A(t,x)=-\frac{q}{2}\int|x-y|\psi^{\dagger}(t,y)\psi(t,y) dy+A_{C}(t,x),
\end{align}
where $A_{C}$ is a potential for the classical source, i.e.,
\begin{align}\label{eq:classical source}
     J^{t}_{C}(t,x)=\partial_{x}E_{C}(t,x)=-\partial_{x}^{2}A_{C}(t,x).
\end{align}
Eq.~\eqref{eq:classcial A} shows that $A(t,x)$ at time $t$ is determined only by the fermion field and the classical source at time $t$ without solving the evolution equation, showing the non-dynamical nature of the gauge field.

The Hamiltonian is obtained by Legendre transforming Eq.~(\ref{eq:Lagrangian QED}). Eliminating $A(t,x)$ using Eq.~(\ref{eq:classcial A}) and discarding the surface integral terms as well as a non-dynamical term proportional to $A_C \partial_x^2 A_C$, one finds
\begin{align}\label{eq QED classical hamiltonian}
     H(t) & = \int dx\left[-i\overline{\psi}\gamma^{x}\partial_{x}\psi+m\overline{\psi}\psi+qA_{C}\psi^{\dagger}\psi\right] \\ & -\frac{q^{2}}{4}\iint dxdy|x-y|\psi^{\dagger}(t,x)\psi(t,x)\psi^{\dagger}(t,y)\psi(t,y). \nonumber
\end{align}

At this point, everything is classical.  The quantum theory in the Schr\"odinger picture is obtained by promoting the fermion field $\psi(t_{0},x)$ at some time $t_{0}$ to a time-independent operator $\widehat{\psi}(x)$ that satisfies the canonical anticommutation relation,
\begin{align}
    \left\{\widehat{\psi}_{\alpha}(x),\widehat{\psi}_{\beta}^{\dagger}(y)\right\}
    & =\delta_{\alpha \beta}\delta(x-y). 
\end{align}
We choose the time $t_0$ at which the Schr\"odinger picture operator is defined to be equal to the time at which the external classical electric field is turned on. To avoid the infinite zero-point energy we normal-order the Hamiltonian as follows,
\begin{equation}\label{eq QED quantum hamiltonian}\begin{aligned}
    & \widehat{H}(t)=\int dx:\!\left[-i\overline{\widehat{\psi}}\gamma^{x}\partial_{x}\widehat{\psi}+m\overline{\widehat{\psi}}\widehat{\psi}+qA_{C}\widehat{\psi}^{\dagger}\widehat{\psi}\right]\!: \\
    & -\frac{q^{2}}{4}\iint dxdy:\!\widehat{\psi}^{\dagger}(x)\widehat{\psi}(x)\!:|x-y|:\!\widehat{\psi}^{\dagger}(y)\widehat{\psi}(y)\!:.
\end{aligned}\end{equation}
This particular normal-ordering makes the Hamiltonian \eqref{eq QED quantum hamiltonian} differ from the bare Hamiltonian (Eq.~\eqref{eq QED classical hamiltonian} with $\psi(t_{0},x)\to \widehat{\psi}(x)$) by
\begin{align}
    \widehat{H}=\widehat{H}_{\rm bare} + \mathcal{C}_1 + \widehat{Q} \mathcal{C}_2,
\end{align}
where $\mathcal{C}_1$ and $\mathcal{C}_2$ are infinite $c$-numbers and $\widehat{Q}$ is the charge operator,
\begin{align}\label{eq:charge operator}
\widehat{Q}=\int:\!\widehat{\psi}^{\dagger}(x)\widehat{\psi}(x)\!:dx.
\end{align}
Thus the Hamiltonians describe the same physics provided that we restrict to states of zero charge.  This assumption is consistent with time evolution since $\hat{Q}$ commutes with the Hamiltonian.  We will assume zero total charge for the remainder of the paper.

Keep in mind that we are in the Schr\"odinger picture even though the Hamiltonian $\widehat{H}$ depends on time. This is simply due to the fact that the external classical field is time-dependent, and the energy of the system is not generally conserved. 

Now, we are ready to bosonize this Hamiltonian.  The boson field $\widehat{\phi}$ and its canonical momentum $\widehat{\Pi}$ satisfy the canonical communtation relations,
\begin{align}
    \label{eq:commu phi,pi}
    \left[\widehat{\phi}(x),\widehat{\Pi}(y)\right]&=i\delta(x-y),
    \\
    \label{eq:commu phi,phi}
    \left[\widehat{\phi}(x),\widehat{\phi}(y)\right]&=0,
    \\
    \label{eq:commu pi,pi}
    \left[\widehat{\Pi}(x),\widehat{\Pi}(y)\right]&=0.
\end{align}
and the bosonization identities are\footnote{The correspondence between the fermi and boson theories has been extensively studied, and an overwhelming amount of evidence suggests that the two theories are equivalent when these substitutions are made in the Hamiltonian (e.g. \cite{Coleman1975apra,Mandelstam1975maya,Haldane:1981zza,Witten:1983ar,Naon1985apr,vonDelft:1998pk,Senechal:1999us,Shankar:2017zag,Justine2021}).}
(e.g. \cite{Coleman.Jackiw.ea1975sep,Coleman1976sep,Smilga:1992hx})
\begin{align}
    \label{eq:fermi=boson kinetic}
    :\!-i\widehat{\psi}\gamma^{x}\partial_{x}\widehat{\psi}\!:
    &\longleftrightarrow N_{\mu}\left\{\frac{1}{2}\widehat{\Pi}^{2}+\frac{1}{2}(\partial_{x}\widehat{\phi})^{2}\right\},
    \\
    \label{eq:fermi=boson jt}
    :\!\overline{\widehat{\psi}}\gamma^{t}\widehat{\psi}\!:&\longleftrightarrow \frac{1}{\sqrt{\pi}}\partial_{x}\widehat{\phi},
    \\
    \label{eq:fermi=boson jx}
    :\!\overline{\widehat{\psi}}\gamma^{x}\widehat{\psi}\!:&\longleftrightarrow-\frac{1}{\sqrt{\pi}}\widehat{\Pi},
    \\
     \label{eq:fermi=boson mass}
    :\!\overline{\widehat{\psi}}\widehat{\psi}\!: &\longleftrightarrow-\frac{e^{\gamma}\mu}{2\pi}N_{\mu}\left\{\cos{\sqrt{4\pi}\widehat{\phi}}\right\}.
\end{align}
The symbol $N_\mu$ denotes normal ordering with respect to an arbitrary mass scale $\mu$.  This procedure is defined by mode expanding $\widehat{\phi}(x)$ and $\widehat{\Pi}(x)$ as 
\begin{align}
    \label{eq:phi expansion}
    \widehat{\phi}(x)&=\int_{-\infty}^{\infty}\frac{dk}{2\pi\sqrt{2\omega_{k,\mu}}}\left(\widehat{a}_{k,\mu}e^{ikx}+\widehat{a}^{\dagger}_{k,\mu}e^{-ikx}\right),
    \\
    \label{eq:Pi expansion}
    \widehat{\Pi}(x)&=
    -i\int_{-\infty}^{\infty}\frac{dk}{2\pi}\sqrt{\frac{\omega_{k,\mu}}{2}}
    \left(\widehat{a}_{k,\mu}e^{ikx}-\widehat{a}^{\dagger}_{k,\mu}e^{-ikx}\right),
\end{align}
where
\begin{align}\label{eq:frequency}
    \omega_{k,\mu}=\sqrt{k^{2}+\mu^{2}}.
\end{align}
Then, $N_{\mu}\left\{\cdots\right\}$ means to rearrange the order of operators by placing all $\widehat{a}_{k,\mu}$ to the right of all $\widehat{a}^{\dagger}_{k,\mu}$.  

The choice of $\mu$ is fully arbitrary.  A change in $\mu$ shifts the kinetic terms \eqref{eq:fermi=boson kinetic} by an infinite c-number and leaves the cosine term \eqref{eq:fermi=boson mass} invariant \cite{Coleman1976sep}, so that the dynamics are unmodified.  At the level of the substitution rules, there is no natural choice of $\mu$.  However, the original theory has mass scales of $m$ and $q$.  Furthermore, the last term in \eqref{eq QED quantum hamiltonian} bosonizes to $\int \tfrac{1}{2\pi}q^2 \widehat{\phi}^2$ (dropping boundary terms) indicating the presence of a mass
\begin{align}
    \label{eq:M=q/sqrt{pi}}
    M \equiv \frac{q}{\sqrt{\pi}}.
\end{align}
Since we will be treating the theory perturbatively in the fermion mass $m$, this ``meson mass'' $M$ is the most natural scale for normal ordering.  Adopting this choice, the bosonized Hamilton becomes
\begin{align}
    \label{eq:Hamiltonian QED boson}
    \widehat{H}(t)=\int dxN_{M}\left\{\frac{1}{2}\widehat{\Pi}(x)^{2}+\frac{1}{2}\left(\partial_{x}\widehat{\phi}(x)\right)^{2}
    +\frac{1}{2}M^{2}\widehat{\phi}(x)^{2}\right.\notag
    \\
    \left.
    +\frac{q}{\sqrt{\pi}}E_{C}(t,x)\widehat{\phi}(x)
    -\frac{e^{\gamma}mM}{2\pi}\cos{\sqrt{4\pi}\widehat{\phi}(x)}
    \right\}.
\end{align}
Notice that at $m=0$, $M$ has a clear physical meaning as the mass of  $\widehat{\phi}(x)$ particle.  This mass is also present in the cosine self-interaction term.  On the other hand, the coefficient $q/\sqrt{\pi}$ in front of $E_{C}(t,x)$ functions as an electric coupling, so we do not express it in terms of $M$, even though it has the same numerical value.

It is worth emphasizing that the choice $\mu=M$ of normal ordering mass is purely for convenience.  This choice allows us to separate the Hamiltonian (\ref{eq:Hamiltonian QED boson}) into a free part, which evolves as a free field of mass $M$, and an interaction part that only contains the cosine term.  If we had chosen a different value for $\mu$, we would need to evolve the free part with mass $\mu$,  and the interaction Hamiltonian would include terms quadratic in $\widehat{\phi}$, unnecessarily complicating the calculations. 

The Schr\"odinger-picture field operator $\widehat{A}(t,x)$ is obtained by replacing $\psi(t,x)$ with $\widehat{\psi}(x)$ in Eq.~(\ref{eq:classcial A}).  The electric field operator $\widehat{E}(t,x)$ is then given by
\begin{align}
    \label{eq:quantum electric field}
    \widehat{E}(t,x)=-\partial_{x}\widehat{A}(t,x)=\frac{q}{\sqrt{\pi}}\widehat{\phi}(x)+E_{C}(t,x).
\end{align}
Although the operators $\widehat{A}$ and $\widehat{E}$ are defined in the Schr\"odinger picture, they inherit explicit time-dependence from the classical field. Eq.~(\ref{eq:quantum electric field}) gives a nice physical meaning to the boson field $\widehat{\phi}(x)$: it is the quantum part of the electric field.

Before proceeding, it is worth noting the transformation properties of Eq.~\eqref{eq:quantum electric field}. In 1+1 dimensions, the electric field $E=F_{10}$ is invariant under Lorentz boosts but flips sign under parity transformations, characterizing it as a pseudoscalar.  The bosonized field $\widehat{\phi}(x)$ also transforms as a pseudoscalar according to Eq.~\eqref{eq:fermi=boson jt}.


\section{Vacuum expectation value}\label{sec:expectation value}
We consider the case where the electric field is zero at an early time and turned on at $t=t_{0}$.\footnote{In QED$_{2}$, a constant $E_{C}$ is sometimes referred to as a ``$\theta$ term'' (e.g.~\cite{Coleman1976sep}). Here, setting $E_{C}=0$ before $t=t_{0}$ indicates that we consider the case of external fields where $\theta=0$.} Before the external field is turned on, we assume that the state is in the vacuum state $\ket{\Omega}$ defined to be the lowest energy eigenvector of the Hamiltonian (\ref{eq:Hamiltonian QED boson}) at $t<t_{0}$.  Since the bosonized Hamiltonian (\ref{eq:Hamiltonian QED boson}) is considered to be equivalent to the original fermion Hamiltonian (\ref{eq QED quantum hamiltonian}), the vacuum state $\ket{\Omega}$ defined on the boson side should also be the vacuum state on the fermion side. 
After turning on the field ($t>t_{0}$), $\ket{\Omega}$ evolves to a state $\ket{t;\Omega}$ according to the Schr\"odinger equation:
\begin{align}
    i\frac{d}{dt}\ket{t;\Omega}=\widehat{H}(t)\ket{t;\Omega}
\end{align}
Since the $\widehat{\phi}(x)$ is the scaled electric field due to Eq.~(\ref{eq:quantum electric field}), the expectation value of $\widehat{\phi}(x)$ evaluated on $\ket{t;\Omega}$ can capture the time evolution of the vacuum as a response to the external field; thus, our goal is to compute 
\begin{align}
    \braket{\phi}(t,x) \equiv \bra{t;\Omega}\widehat{\phi}(x)\ket{t;\Omega}.
\end{align}

\subsection{Interaction picture}
 
We now split the Hamiltonian (\ref{eq:Hamiltonian QED boson}) into a free part (which includes the classical source) and an interaction part (the cosine term). To explicitly write down the vacuum $\ket{\Omega}$, we introduce an adiabatic switching function that turns off the interaction at early times.

We define the total Hamiltonian $\widehat{H}_{T}(t)$ by 
\begin{align}
    \label{eq:total Hamiltonian}
    \widehat{H}_{T}(t)=&\widehat{H}_{0}(t)+\widehat{H}_{\rm int}(t)
\end{align}
where
\begin{align}
    \widehat{H}_{0}(t)=\widehat{H}_{\rm  ff}+\widehat{H}_{E}(t)
\end{align}
and each piece is given by 
\begin{align}
    \label{eq:free-free Hamiltonian}
    &\widehat{H}_{\rm ff}=\int \! dxN_{M}\left[
    \frac{1}{2}\widehat{\Pi}(x)^{2}+\frac{1}{2}(\partial_{x}\widehat{\phi}(x))^{2}
    +\frac{1}{2}M^{2}\widehat{\phi}(x)^{2}
    \right],
    \\
    \label{eq:E Hamiltonian}
    &\widehat{H}_{E}(t)=\int\! dx
    \frac{q}{\sqrt{\pi}}E_{C}(t,x)\widehat{\phi}(x),
    \\
    \label{eq:int Hamiltonian}
    &\widehat{H}_{\rm int}(t)=\int \!dxN_{M}\left[
    -\frac{e^{\gamma}mf(t)M}{2\pi}\cos{\sqrt{4\pi}\widehat{\phi}(x)}
    \right].
\end{align}
$\widehat{H}_{\rm ff}$ represents the ``free-free" part of the Hamiltonian in the sense that this Hamiltonian describes a free massive particle with mass $M$ with no interaction or electric field. 
The function $f(t)$ in $\widehat{H}_{\rm int}(t)$ is chosen to vanish at an early time and to reach $f=1$ before $t=t_{0}$ (the time at which the external classical field is turned on).  In addition, we demand that the transition of $f(t)$ from 0 to 1 is infinitesimally slow.  This procedure allows us to write down the expression of $\ket{\Omega}$ in terms of operators and states defined in the interaction picture as explained below.

In the infinite past, there is no external electric field and the interaction term is turned off due to the $f(t)$; thus, the total Hamiltonian is given by
\begin{align}\label{eq: Hinitial}
    \widehat{H}_{T}(-\infty)=\widehat{H}_{\rm ff}.
\end{align}
The free-free Hamiltonian $\widehat{H}_{\rm ff}$ defines a vacuum state $\ket{0,M}$ in the usual way, i.e. by 
\begin{align}
\widehat{a}_{k,M}\ket{0,M}=0
\end{align}
for all $k$ with $\widehat{a}_{k,M}$ defined in 
\eqref{eq:phi expansion} and \eqref{eq:Pi expansion}.  
This provides the initial value fo the true vacuum state according to \eqref{eq: Hinitial}.  
Since $f(t)$ turns on very slowly, we may evolve the vacuum using the adiabatic approximation.  In particular, the true vacuum $\ket{\Omega}$ can be approximated by evolving $\ket{0,M}$ from the infinite past to time $t_{0}$ using the Hamiltonian (\ref{eq:Hamiltonian QED boson}),
\begin{align}
    \label{eq:define |Omega>}
    \ket{\Omega}&\simeq \widehat{U}_{T}(t_{0},-\infty)\ket{0,M}.
\end{align}
Here, $\simeq$ means that two expressions differ only by a phase, and $\widehat{U}_{T}(t,t')$ is the evolution operator for $\widehat{H}_T$, defined by Eq.~(\ref{eq:define U(t,t')}). 

The interaction picture is obtained by evolving the operators with the free time-evolution operator $\widehat{U}_{0}(t,t')$ and evolving the state with the interaction evolution operator $\widehat{U}_{\rm int}(t,t')$.  Since our Hamiltonian depends on time, the theory is slightly modified from the most familiar form, and for the reader's convenience, we review the relevant equations in Appendix \ref{app:interactting picutre}.

The interaction picture operators $\widehat{\phi}^{I}(t,x)$ and $\widehat{\Pi}^{I}(t,x)$ are defined by (see Eq.~(\ref{eq:define interaction O})),
\begin{align}
    \label{eq: define phi interaction}
    \widehat{\phi}^{I}(t,x)&=\widehat{U}_{0}^{\dagger}(t,t_{0})\widehat{\phi}(x)\widehat{U}_{0}(t,t_{0}),
    \\
    \label{eq: define pi interaction}
    \widehat{\Pi}^{I}(t,x)&=\widehat{U}_{0}^{\dagger}(t,t_{0})\widehat{\Pi}(x)\widehat{U}_{0}(t,t_{0}).
\end{align}
The above definition ensures the following conditions
\begin{align}
    \label{eq: phi interaction ini}
     \widehat{\phi}^{I}(t_{0},x)&= \widehat{\phi}(x),
     \\
     \label{eq: Pi interaction ini}
     \widehat{\Pi}^{I}(t_{0},x)&= \widehat{\Pi}(x).
\end{align}
Using Eqs.~(\ref{eq:commu phi,pi})--(\ref{eq:commu pi,pi}), one can show that these operators satisfy the equal-time commutation relation
\begin{align}
    \label{eq:equal time phi pi commu interaction}
    \left[\widehat{\phi}^{I}(t,x),\widehat{\Pi}^{I}(t,y)\right]&=i\delta(x-y),
    \\
    \label{eq:equal time phi phi commu interaction}
    \left[\widehat{\phi}^{I}(t,x),\widehat{\phi}^{I}(t,y)\right]&=0,
    \\
    \label{eq:equal time pi pi commu interaction}
    \left[\widehat{\Pi}^{I}(t,x),\widehat{\Pi}^{I}(t,y)\right]&=0.
\end{align}
Then, from Eq.~(\ref{eq:heisenberg O interaction}) and using Eqs.~(\ref{eq:equal time phi pi commu interaction})--(\ref{eq:equal time pi pi commu interaction}), we have the following operator equations for $\widehat{\phi}^{I}(t,x)$ and $\widehat{\Pi}^{I}(t,x)$:
\begin{align}
    \label{eq:phi heisenberg eq}
    \frac{\partial}{\partial t}\widehat{\phi}^{I}(t,x)
    &=-i\left[\widehat{\phi}^{I}(t,x),\widehat{H}^{I}_{0}(t)\right]= \widehat{\Pi}^{I}(t,x),
    \\\notag
    \\
    \label{eq:pi heisenberg eq}
       \frac{\partial}{\partial t}\widehat{\Pi}^{I}(t,x)
       &=-i\left[\widehat{\Pi}^{I}(t,x),\widehat{H}^{I}_{0}(t)\right]
       \\
       &=\partial_{x}^{2}\widehat{\phi}^{I}(t,x)-\mu^{2}\widehat{\phi}^{I}(t,x)-\frac{q}{\sqrt{\pi}}E_{C}(t,x).\notag
\end{align}
The explicit solutions to these equations are given by
\begin{widetext}
\begin{align}
    \label{eq:op sol phi interaction}
    \widehat{\phi}^{I}(t,x)&=\int_{-\infty}^{\infty}\frac{dk}{2\pi\sqrt{2\omega_{k,M}}}\left(\widehat{a}_{k,M}e^{-i\omega_{k,M}(t-t_{0})+ikx}+\widehat{a}^{\dagger}_{k,M}e^{i\omega_{k,M}(t-t_{0})-ikx}\right)+\phi_{\rm cl}(t,x),
    \\
    \label{eq:op sol pi interaction}
    \widehat{\Pi}^{I}(t,x)&=-i\int_{-\infty}^{\infty}\frac{dk}{2\pi}\sqrt{\frac{\omega_{k,M}}{2}}\left(\widehat{a}_{k,M}e^{-i\omega_{k,M}(t-t_{0})+ikx}-\widehat{a}^{\dagger}_{k,M}e^{i\omega_{k,M}(t-t_{0})-ikx}\right)+\partial_{t}\phi_{\rm cl}(t,x),
\end{align}
\end{widetext}
where $\omega_{k,M}$ is defined by Eq.(\ref{eq:frequency}). The classical part $\phi_{\rm cl}(t,x)$ satisfies the Klein-Gordon equation
\begin{align}\label{eq: KG cl}
    \left(\partial_{t}^{2}-\partial_{x}^{2}+M^{2}\right)\phi_{\rm cl}(t,x)=-\frac{q}{\sqrt{\pi}}E_{C}(t,x),
\end{align}
with the condition 
\begin{align}
\phi_{\rm cl}(t,x)=0=\partial_{t}\phi_{\rm cl}(t,x) \quad \textrm{for } t\leq t_{0}.
\end{align}
This condition is a consequence of the fact that $\widehat{\phi}^{I}(t,x)$ and $\widehat{\Pi}^{I}(t,x)$ need to satisfy the conditions (\ref{eq: phi interaction ini}) and (\ref{eq: Pi interaction ini}) as well as the fact that $E_{C}(t,x)$ vanishes at $t\leq t_{0}$.  Equivalently, $\phi_{\rm cl}$ is the retarded solution to Eq.~\eqref{eq: KG cl}. 

The free evolution operator before the electric field turns on only changes $\ket{0,M}$ by a phase, i.e., 
\begin{align}
    \widehat{U}_{0}(t_{0},-\infty)\ket{0,M}\simeq \ket{0,M}. 
\end{align}
The expression of the state $\ket{t;\Omega}$ is then 
\begin{align}
    \label{eq: formaula for |t,Omega>}
    \ket{t;\Omega}&=\widehat{U}_{T}(t,t_{0})\ket{\Omega} \notag\\
    &\simeq\widehat{U}_{0}(t,t_{0})\widehat{U}_{\rm int}(t,-\infty)\ket{0,M}.
\end{align}
using \eqref{eq:define |Omega>} and \eqref{eq:define Uint(t,t')}.  Then, the expression of $\braket{\phi}=\bra{t;\Omega}\widehat{\phi}(x)\ket{t;\Omega}$ can be written in the interaction picture by 
\begin{align}
    \label{eq:<phi> int picture}
    \braket{\phi}
    &=\bra{0,M}\widehat{U}_{\rm int}^{\dagger}(t,-\infty)\widehat{\phi}^{I}(t,x)\widehat{U}_{\rm int}(t,-\infty)\ket{0,M}.
\end{align}
with $\widehat{U}_{\rm int}(t,t')$ given by Eq.~(\ref{eq:U int}). 

To proceed further, an important observation is that the normal ordering $N_{M}\left\{\cdots\right\}$ and the operation to obtain the interaction picture operators commute, i.e., for a function $F(\widehat{\phi})$ of polynomials of $\widehat{\phi}$ (not necessarily evaluated at the same spatial points), we have
\begin{align}
    \label{eq: NM commutes with U0}
    \widehat{U}_{0}^{\dagger}(t,t_{0})N_{M}\left\{F(\widehat{\phi})\right\}\widehat{U}_{0}(t,t_{0})
    =N_{M}\left\{F(\widehat{\phi}^{I})\right\}.
\end{align}
This allows us to write $\widehat{U}_{\rm int}(t,t')$ fully in terms of $\widehat{\phi}^{I}(t,x)$ while keeping the structure of normal ordering intact.  In particular, from (\ref{eq:Uint dyson series}) and (\ref{eq:int Hamiltonian}) the Dyson series solution is
\begin{align}
    \label{eq:U int}
    &\widehat{U}_{\rm int}(t,t')=\notag
    \\
    &T\exp{\left(i\!\int^{t}_{t'}ds\int\! dx N_{M}\!\left\{\frac{e^{\gamma} m Mf(s)}{2\pi}\cos{\sqrt{4\pi}\widehat{\phi}^{I}(s,x)}\right\}\right)},
\end{align}
where $T$ denotes time ordering. 
Note that Eq.~(\ref{eq:U int}) is a shorthand notation for a Taylor series in  $m$ and will provide an asymptotic approximation as $m\to 0$ as long as each term gives a finite expression.
We carry out the expansion to the first order in $m$ and find a finite result for $\braket{\phi}$.

\subsection{\texorpdfstring{Calculation of $\braket{\phi}$}{Calculation of phi}}

It is helpful to review a few facts before computing the field expectation value.  First, we have
the normal-ordering identity
\begin{align}
    \left[e^{\widehat{A}},\widehat{B}\right]=&\left[\widehat{A},\widehat{B}\right]e^{\widehat{A}},
\end{align}
which holds for any operators $\widehat{A}$ and $\widehat{B}$ whose commutator is a c-number.  This in particular implies that
\begin{align}
    \label{eq:cos phi commu interaction}
    &\left[ N_{M}\left\{\cos{\sqrt{4\pi}\widehat{\phi}^{I}(t_{1},y_{1})}\right\},\widehat{\phi}^{I}(t_{2},x_{2})\right]
    \\
     &=-\sqrt{4\pi}\left[\widehat{\phi}^{I}(t_{1},y_{1}),\widehat{\phi}^{I}(t_{2},y_{2})\right] N_{M}\left\{\sin{\sqrt{4\pi}\widehat{\phi}^{I}(t_{1},y_{1})}\right\}. \notag
\end{align}

Another useful fact is that for any composite operators $F(\widehat{\phi}^{I}(t,x),\widehat{\Pi}^{I}(t,x))$ formed from polynomials of $\widehat{\phi}^{I}(t,x)$ and $\widehat{\Pi}^{I}(t,x)$, we have 
\begin{align}
    \label{eq:op to cl relatio}
    &\bra{0,M}N_{M}\left\{F\left(\widehat{\phi}^{I}(t,x),\widehat{\Pi}^{I}(t,x)\right)\right\}\ket{0,M}\notag
    \\
    &\hspace{3cm}=
    F\left(\phi_{\rm cl}(t,x),\partial_{t}\phi_{\rm cl}(t,x)\right).
\end{align}
To check, notice that the states on both sides annihilate all the operator parts of $\widehat{\phi}^{I}(t,x)$ and $\widehat{\Pi}^{I}(t,x)$, leaving only the classical parts. 

Finally, we note that the retarded Green function for Klein-Gordon equation may be expressed in terms of the interaction-picture fields as (e.g. \cite{Peskin:1995ev}) 
\begin{align}
    &G_{\rm ret}(\Delta t,\Delta x) = i\theta(\Delta t)\left[\widehat{\phi}^{I}(t,x),\widehat{\phi}^{I}(t',y)\right].\label{eq: Gret as commutator}
\end{align}
where $\Delta t=t-t'$ and $\Delta x=x-y$ and $\theta(t) $ is the step function.  The retarded green function satisfies  
\begin{align}
    \left(\partial_{t}^{2}-\partial_{x}^{2}+M^{2}\right)G_{\rm ret}(t,x)=\delta(t)\delta(x).
\end{align}
and is explicitly expressed as (e.g. \cite{Dvornikov:2004xh,Finster.Paganini2023feb}) 
\begin{align}
    & G_{\rm ret}(\Delta t,\Delta x)=\frac{\theta(\Delta t)}{\pi}\int^{\infty}_{-\infty}\!\frac{dk}{2\pi \omega_{k,M}}\sin{(\omega_{k,M}\Delta t)}e^{ik\Delta x}\notag
    \\
    &=\frac{1}{2}\theta\left(\Delta t\right)\theta\left(\Delta t^{2}-\Delta x^{2}\right)J_{0}\left(M\sqrt{\Delta t^{2}-\Delta x^{2}}\right).
\end{align}

With the ingredients assembled, the calculation of $\braket{\phi}$ is straightforward,
\begin{widetext}
\begin{align}
    \label{eq:expectation value phi}
    \braket{\phi}&=\bra{0,M}\widehat{\phi}^{I}(t,x)\ket{0,M}+\int_{-\infty}^{t}dt'\int\!dy
    \left(-i\frac{e^{\gamma}mMf(t')}{2\pi\sqrt{\pi}}\right)
    \bra{0,M}\left[N_{M}\left\{\cos{\sqrt{4\pi}\widehat{\phi}^{I}(t',y)}\right\},\widehat{\phi}^{I}(t,x)\right]\ket{0,M}\notag
    \\
    &=\phi_{\rm cl}(t,x)+
    \int_{-\infty}^{t}dt'\int\!dy
    \left(i\frac{e^{\gamma}mMf(t')}{2\pi\sqrt{\pi}}\right)
    \sqrt{4\pi}\left[\widehat{\phi}^{I}(t',y),\widehat{\phi}^{I}(t,x)\right]
    \bra{0,M}N_{M}\left\{\sin{\sqrt{4\pi}\widehat{\phi}^{I}(t',y)}\right\}\ket{0,M}\notag
    \\
    &=
    \phi_{\rm cl}(t,x)+
     \int_{-\infty}^{\infty}dt'\int\!dy
    \left(-\frac{e^{\gamma}mq}{\pi}\right)\sin{\left[\sqrt{4\pi}\phi_{\rm cl}(t',y)\right]}
    G_{\rm ret}(t-t',x-y).
\end{align}
\end{widetext}
The first line follows from  \eqref{eq:<phi> int picture} and \eqref{eq:U int}, while the second and third lines use \eqref{eq:cos phi commu interaction} and \eqref{eq: Gret as commutator}, respectively.  In the third line we are able to set $f(t')=1$ since $\phi_{\rm cl}(t,x)=0$ at $t<t_{0}$.  Eq.~(\ref{eq:expectation value phi}) is the causal perturbative solution to Eq.~(\ref{eq:PDE for <phi>}), justifying the claim that $\braket{\phi}$ obeys this equation to $O(m)$.

\subsection{\texorpdfstring{Validity of the $m$ expansion}{Validity of the m expansion}}

The perturbative expression \eqref{eq:expectation value phi} for $\braket{\phi}$ is likely to provide a good approximation as long as the perturbation is numerically small compared to the background term $\phi_{\rm cl}$.
The comparison may be made more easily at the level of the differential equation (\ref{eq:PDE for <phi>}).  For this purpose we will also restore factors of $\hbar$, noting that $\phi$ is dimensionless, while $q$ and $E_{C}$ have the same units and  $qE_{C}$ has units of force.  Eq.~(\ref{eq:PDE for <phi>}) then becomes
\begin{align}
    \label{eq:PDE for <phi> hbar}
    &\left(\partial_{t}^{2}-\partial_{x}^{2}+\frac{q^{2}}{\hbar\pi}\right)\braket{\phi}+\frac{e^{\gamma}mq}{\hbar^{3/2}\pi}\sin\left(\sqrt{4\pi}\braket{\phi}\right)=-\frac{q}{\hbar\sqrt{\pi}}E_{C}.
\end{align}
The non-analyticity in $\hbar$ indicates that effects predicted by Eq.~(\ref{eq:PDE for <phi> hbar}) cannot be captured in the limit of $\hbar\to 0$.

In order for the mass expansion to be valid, we require that the sine term of Eq.~(\ref{eq:PDE for <phi> hbar}) is much smaller than at least one other term.  There are two cases where this assumption is justified.

The first case is the \textit{strong field limit}.  If the field has typical strength $E_C\sim\mathscr{E}$, then the condition $\mathscr{E}\gg m/\hbar^{1/2}$ guarantees that the sine term in Eq.~\eqref{eq:PDE for <phi> hbar} is small compared to the electric field.  However, since $E_{C}$ is  ``turned on'' from zero starting at time $t=t_{0}$, there may be times when it is not small compared to the sine term.  We expect that if the transition is sufficiently rapid then the approximation will remain valid, and we may see this in the equations as follows.  If $\mathcal{T}$ denotes the timescale for $E_C$ to reach $\mathscr{E}$, then we expect $\mathcal{T}$ to set the timescale of variation for $\braket{\phi}$ during this period.   Using $\partial_t^2\braket{\phi}\sim \braket{\phi}/\mathcal{T}^2$, we find the condition $e^\gamma mq/\hbar^{3/2} \ll \pi /\mathcal{T}^{2}$ for the sine term to be small compared to $\partial_t^2\braket{\phi}$. Thus a regime of validity for our approximation is 
\begin{align}
    \label{eq:strong field}
    \mathscr{E} \gg \frac{m}{\hbar^{1/2}}, \quad \mathcal{T} \ll \sqrt{\frac{\hbar^{3/2}}{mq}}.
\end{align}
The second case where small $m$ treatment is justified is \textit{fermion strong coupling limit}.  We see directly in Eq.~\eqref{eq:PDE for <phi>} that the sine term is small compared to the mass term $q^2\braket{\phi}/(\hbar\pi)$ provided 
\begin{align}
    \label{eq:strong coupling}
    q\gg \frac{m}{\hbar^{1/2}}. 
\end{align}
This regime brings out the strong-weak nature of the duality; the strong coupling $q \gg m/\hbar^{1/2}$ on the fermion side translates to weak coupling on the boson side, enabling perturbative calculations like ours.

Notice that the conditions of validity (\ref{eq:strong field}) and (\ref{eq:strong coupling}) cannot be satisfied if one takes $\hbar\to0$.  This is the sense in which our result may be considered ``fully quantum''.


\subsection{Conservation of energy}\label{sec:conserved quantity}

Before proceeding, we note that Eq.~\eqref{eq:PDE for <phi>} admits a conserved energy when the external field $E_C$ is time-independent.  We express this quantity as
\begin{align}
    \label{eq: total energy}
\mathcal{E}=\mathcal{E}_{\text{particle}}+\mathcal{E}_{\text{field}},
\end{align}
where
\begin{align}
    \label{eq: particle energy}
    \mathcal{E}_{\text{particle}}
    &=
    \int\left[\frac{1}{2}\left(\partial_{t}\braket{\phi}\right)^{2}+\frac{1}{2}\left(\partial_{x}\braket{\phi}\right)^{2}\right.\notag
    \\
    & \qquad \qquad \left.-
    \frac{e^{\gamma}mM}{2\pi}\cos{\sqrt{4\pi}\braket{\phi}}\right]dx,
\end{align}
and
\begin{align}
    \label{eq: field energy}
    \mathcal{E}_{\text{field}}
    &= \int \frac{1}{2}\braket{E}^2dx = 
     \int\frac{1}{2}\left(E_{C}+\frac{q}{\sqrt{\pi}}\braket{\phi}\right)^{2}dx
\end{align}
noting Eq.~\eqref{eq:<E>}.  Notice that $\mathcal{E}_{\rm field}$ is just the ``classical'' field energy associated with the expected value $\braket{E}$ of the electric field; likewise, we may think of  $\mathcal{E}_{\rm particle}$ as a fictitious classical matter field energy.  The conservation of $\mathcal{E}$ represents an exchange between field and particle degrees of freedom.

In appendix~\ref{app:energy expectation value} we show that after adding back in the non-dynamical term $\tfrac{1}{2}E_C^2$ to \eqref{eq:Hamiltonian QED boson}, the expectation value of the Hamiltonian is equal to $\mathcal{E}$ to first order in $m$, i.e.
\begin{align}
    \label{eq:<H>=E}
    \braket{\widehat{H}}=\mathcal{E}.
\end{align}
This establishes the fact that $\mathcal{E}$ is truly the energy of the system. 
However, this simplicity does not extend to the division into particle and field: the ``energies'' $\mathcal{E}_{\rm particle}$ and $\mathcal{E}_{\rm  field}$ are not equal to the expectation values of their quantum counterparts.  


\section{Capacitor breakdown revisited}\label{sec: capacitor breakdown}
As shown in \cite{Chu.Vachaspati2010apr,Gold.McGady.ea2021oct,Pham.Zager.ea2025apr}, bosonized 1+1D QED can be used to study the breakdown of a capacitor when its electric field is strong. For the massless case, Ref.~\cite{Chu.Vachaspati2010apr} considered this problem by placing two oppositely charged plates at a separate distance and letting the field evolve according to Eq.~(\ref{eq:PDE for <phi>}) with zero fermion mass. The massive case was considered in Ref.~\cite{Gold.McGady.ea2021oct} by solving Eq.~(\ref{eq:PDE for <phi>}) assuming a large boson occupation to justify the classical treatment.  Ref.~\cite{Pham.Zager.ea2025apr} developed a method to accurately track the long-term evolution of the field for both massless and massive cases.

Our fully quantum calculation requires the external electric field to vanish in the asymptotic past.  In order to ``turn on'' the electric field in a manner compatible with local conservation of charge, we take our capacitor plates to begin at the same spatial location and then separate.  In particular, the plates start at $x=0$ and and move in opposite direction with velocity $v<1$ until they reach a final separation of $2L$.  This can be modeled by setting the background field to be
\begin{align}
    \label{eq: moving plates}
    E_{C}=\begin{cases} 0, & t<0, \\E_{0}\theta(vt-x)\theta(vt+x), & 0<t < L/v \\
    E_0 \theta(L+x)\theta(L-x) & t > L/v,
    \end{cases}
\end{align}
(We set the time that the electric field turns on to be $t_0=0$.) The charge density associated with this electric field is
\begin{align}
    \label{eq: classical charge}
    \rho_{C}&=\partial_{x}E_{C}\notag
    \\
    &=\begin{cases} 0, & t<0, \\-E_{0}\delta(vt-x)+E_0 \delta(vt+x), & 0<t < L/v \\
    -E_{0}\delta(L-x)+E_0 \delta(L+x) & t > L/v.
    \end{cases}
\end{align}
Thus, the background profile (\ref{eq: moving plates}) corresponds to a positively charged plate moving to the left until it reaches $x=-L$, together with a negatively charged plate moving to the right until it reaches $x=+L$.

The previous work \cite{Chu.Vachaspati2010apr, Gold.McGady.ea2021oct,Pham.Zager.ea2025apr} did not impose any boundary conditions at the capacitor plates. The field can then become non-zero outside, which may be interpreted as allowing the created particles passing through the plates.  Another option we consider is to impose $\braket{\phi}=0$ at the plates, which can be thought of as causing fields and particles to reflect off the plates, as if they are impenetrable mirrors.  
In each case, we solved Eq.~(\ref{eq:PDE for <phi>}) numerically. Since $q$ has the dimension of mass as we can see in Eq.~\eqref{eq:M=q/sqrt{pi}}, we use it as a reference and set other parameters as $m/q=0.1,E_{0}/q=1,v=0.8, qL=20$. 
We use a standard fourth-order Runge-Kutta method to solve the PDE (\ref{eq:PDE for <phi>}).

\begin{figure}[t]
\centering
    \includegraphics[keepaspectratio, scale=0.27]{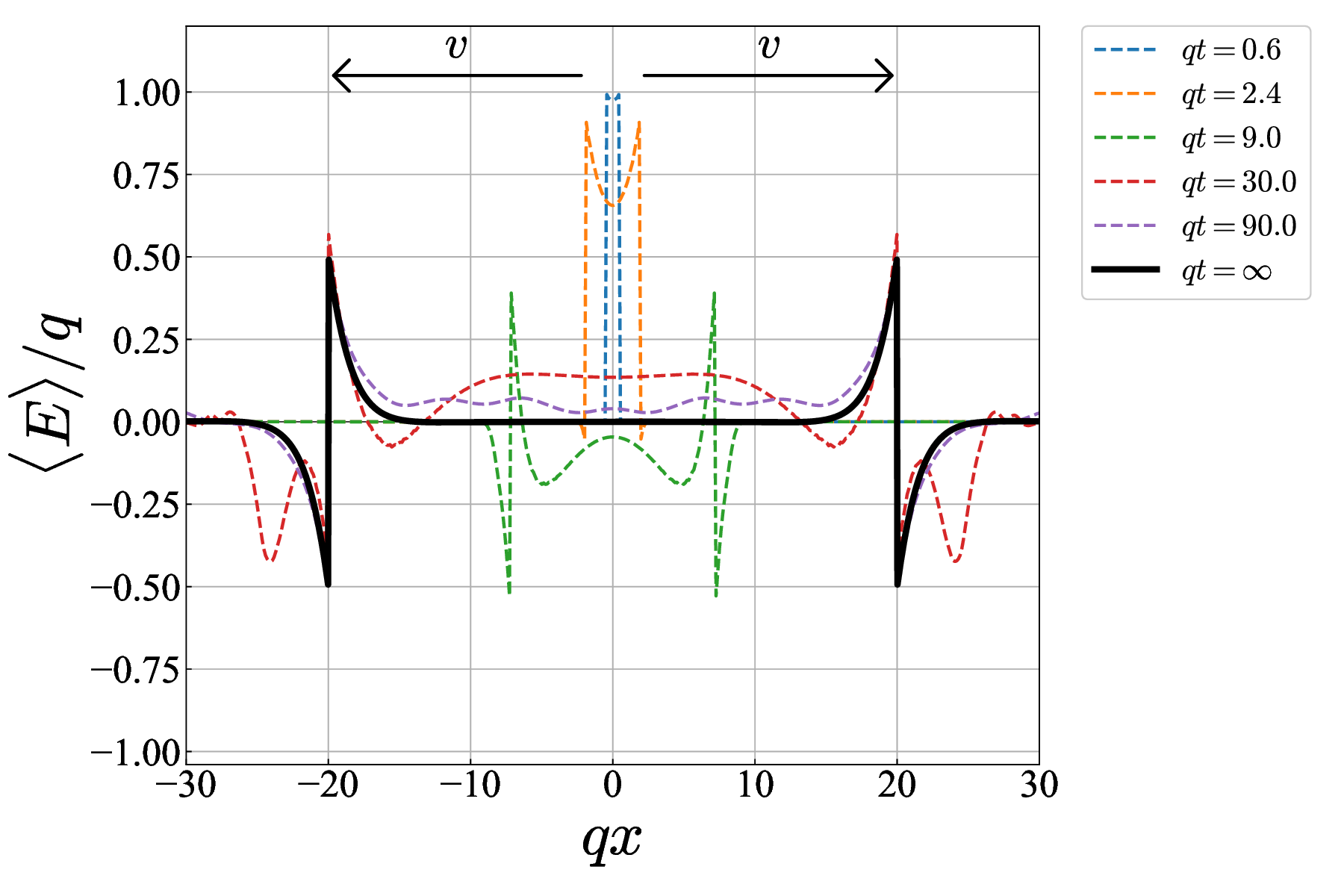}
  \caption{
  \justifying
  The expectation value of the electric field $\braket{E}=E_{C}+q\braket{\phi}/\sqrt{\pi}$.
  Two charged plates start at $x=0$ and move in opposite directions with velocity $v=0.8$. The plates are stopped when they reach $qx=\pm20$. The electric field eventually reaches a static state as all the dynamical part spreads out to infinity. The static solution (black line) shows exponential screening of the plates from the both sides.  
  We set $m/q=0.1,E_{0}/q=1.0$.
  \label{fig: non reflective plates}}
\end{figure}

Fig.~\ref{fig: non reflective plates} shows the case of the penetrable plates. In this case, the time dependence eventually ceases as the dynamical part of the field spreads out to infinity. The final static state of the electric field can be found by solving Eq.~(\ref{eq:PDE for <phi>}) assuming no time dependence, taking the external electric field to be in its final state, with the capacitor fully assembled. 
By expanding $\braket{\phi}$ perturbatively in $m$ as 
\begin{align}
    \label{eq:static phi}
    \braket{\phi}\approx\phi_{0}+\phi_{1},
\end{align}
and plugging this to Eq.~(\ref{eq:PDE for <phi>}), we find 
\begin{align}
    \label{eq:phi0}
    \phi_0(x)=
\begin{cases}
\displaystyle \frac{E_0}{M}
\left(e^{-M L}\cosh{(Mx)}-1\right), & |x|\le L,\\[8pt]
\displaystyle -\frac{E_0}{M}\sinh{(ML)}e^{-M|x|}, & |x|>L~.
\end{cases}
\end{align}
and
\begin{align} 
    \label{eq:phi1}
\phi_1(x)
= -\frac{e^{\gamma}m}{2\sqrt{\pi}}
\int_{-\infty}^{\infty}
e^{-M|x-y|}
\sin{\sqrt{4\pi}\,\phi_0(y)}\,dy~.
\end{align}
The black line in Fig.~\ref{fig: non reflective plates} shows this analytic solution. 
For the final static state, we can see from Eq.~(\ref{eq:phi0}) that the field decays exponentially away from the plates with a characteristic length scale of $1/M$.  One may say that a ``cloud of particles'' surrounds each plate, screening its electric field.  This reproduces the results of \cite{Chu.Vachaspati2010apr} and derives the first correction in fermion mass.  

\begin{figure}[t]
\centering
    \includegraphics[keepaspectratio, scale=0.27]{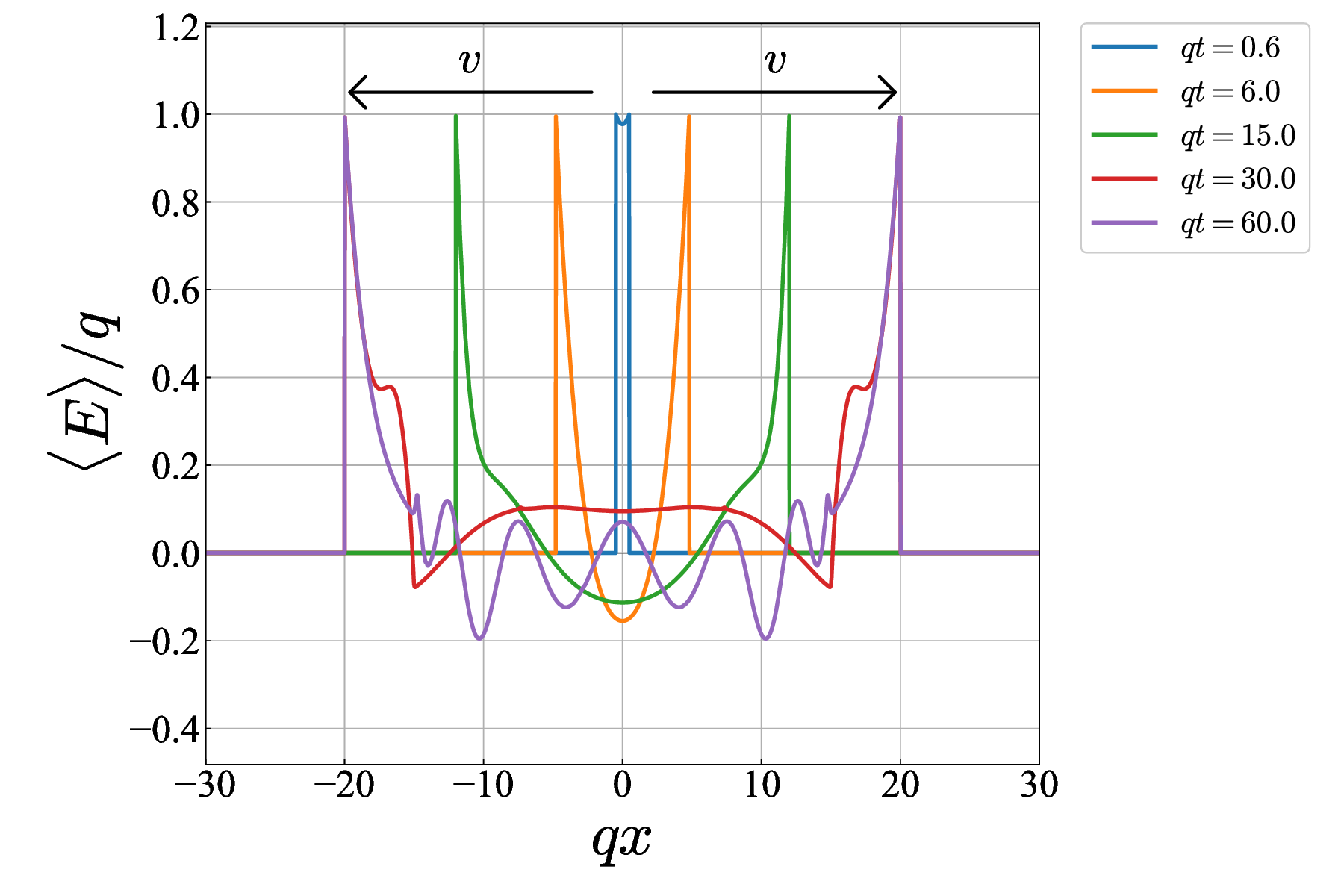}
  \caption{
  \justifying 
  Same as Fig.~\ref{fig: non reflective plates}, except for impenetrable places ($\braket{\phi}$ is forced to vanish outside the plates).  In this case, the electric field inside the capacitor never reaches a static state; the dynamical part of the field bounces back and forth inside the capacitor.
  \label{fig: expanding plates}}
\end{figure}

Fig.~\ref{fig: expanding plates} shows the alternative case of the impenetrable plates, where particles ``bounce off''. In this case, the field keeps oscillating and never settles down to a static state.  This may be attributed to conservation of energy-like quantity (\ref{eq: total energy}).  As one can see Fig.~\ref{fig: expanding plates}, the field is dynamical even after the plates reach the final separation.  This is the same behavior as a classical plasma confined to a box with no dissipation, and the oscillations are known as plasma oscillations.

\section{Plasma frequency for QED}\label{sec:plasma frequency}

A classical plasma consists of point charges of both signs of charge interacting with the self-consistent electromagnetic field.  In this context, local separation between oppositely charged particles results in an electrostatic restoring force that produces oscillations.  The electric field reacts to the motion and also oscillates at the same frequency.  The QED plasma produced by capacitor breakdown also displays these electric field oscillations, although the underlying mechanism may be quite different.  In this section, we derive the frequency of the oscillations and also exhibit an effective potential formulation.

For this purpose, we consider a spatially homogeneous external electric field that turns on at $t=t_{0}$ to the constant value $E_{C}$:
\begin{align}
    E_{C}(t)=E_{C}\theta(t).
\end{align}
We also assume that $\braket{\phi}$ is spatially homogeneous, so that Eq.~(\ref{eq:PDE for <phi>}) becomes an ordinary differential equation,
\begin{align}
    \label{eq:ODE for <phi>}
    \frac{d^{2}}{dt^{2}}\braket{\phi}+M^{2}\braket{\phi}=-\frac{e^{\gamma}mq}{\pi}\sin\left(\sqrt{4\pi}\braket{\phi}\right)-\frac{q}{\sqrt{\pi}}E_{C}.
\end{align}
holding for $t>0$. Since there is no external field for $t<0$, we set $\braket{\phi}=0=\partial_{t}\braket{\phi}$ for $t<0$.

At zeroth order in $m$, Eq.~\eqref{eq:ODE for <phi>} is a simple harmonic oscillator.  The constant external field $E_C$ just shifts the equilibrium position, and the plasma frequency is just $\omega=M$. To find the first correction, we use a standard perturbative approach to a nonlinear ODE known as the Lindstedt–Poincaré method (see \cite{Howes1982,Jordan.Smith2007oct} for details).  We expand the frequency as
\begin{align}\label{eq:omega expand}
    \omega \approx M+\omega_{1},
\end{align}
where $\omega_1$ is order $m$.  We introduce a new time parameter $\tau=\omega t$ and expand $\braket{\phi}$ as
\begin{align}\label{eq:expand phi}
    \braket{\phi}(t) \approx \phi_{0}(\tau)+\phi_{1}(\tau),
\end{align}
where $\phi_{1}(\tau)$ is the order $m$ term. Note that $\tau$ in Eq.~(\ref{eq:expand phi}) is not expanded in $m$. 
By changing the variable from $t$ to $\tau$, Eq.~(\ref{eq:ODE for <phi>}) becomes
\begin{align}\label{eq:ODE for <phi>2}
    \omega^{2}\braket{\phi}''+M^{2}\braket{\phi}=-\frac{e^{\gamma}mq}{\pi}\sin\left(\sqrt{4\pi}\braket{\phi}\right)-\frac{q}{\sqrt{\pi}}E_{C},
\end{align}
where $'$ denotes the $\tau$ derivative. 
Plugging Eqs.~(\ref{eq:omega expand}) and (\ref{eq:expand phi}) into Eq.~(\ref{eq:ODE for <phi>2}) and extracting the zeroth order terms in $m$, we find
\begin{align}
    \label{eq: ODE for phi0}
    M^{2}\phi_{0}''+M^{2}\phi_{0}=-\frac{qE_{C}}{\sqrt{\pi}}.
\end{align}
The solution to this equation satisfying $\phi_{0}(0)=0=\partial_{t}\phi_{0}(0)$ is given by
\begin{align}
    \label{eq:phi0 solution}
    \phi_{0}=\frac{\sqrt{\pi}E_{C}}{q}\left(\cos{\tau}-1\right).
\end{align}
The first order equation from Eq.~(\ref{eq:ODE for <phi>2}) is
\begin{align}
    \label{eq:ODE for phi1}
    \phi_{1}''+\phi_{1}&=-\frac{e^{\gamma}m}{q}\sin{\left(\sqrt{4\pi}\phi_{0}\right)}-\frac{2\sqrt{\pi}\omega_{1}}{q}\phi_{0}''.
\end{align}
Using Eq.~(\ref{eq:phi0 solution}) and Jacobi–Anger expansion
\begin{align}
    \cos{\left(z\cos\theta\right)}&=J_{0}(z)+2\sum_{k=1}^{\infty}(-1)^{k}J_{2k}(z)\cos{(2k\theta)}
    \\
    \sin{\left(z\cos\theta\right)}&=2\sum_{k=0}^{\infty}(-1)^{k}J_{2k+1}(z)\cos{((2k+1)\theta)},
\end{align}
where $J_{k}(x)$ is the $k$th Bessel function of the 1st kind, Eq.~(\ref{eq:ODE for phi1}) becomes
\begin{align}
    \label{eq:ODE for phi 1 2}
    &\phi_{1}''+\phi_{1}=\notag
    \\
    &\left(
    -\frac{2e^{\gamma}m}{q}\cos{\left(\frac{2\pi E_{C}}{q}\right)}J_{1}\left(\frac{2\pi E_{C}}{q}\right)+\frac{2\pi\omega_{1}E_{C}}{q^{2}}
    \right)\notag
    \\
    &\times\cos{\tau}\notag
    \\
    &-\frac{e^{\gamma}m}{q}\sum_{k=1}^{\infty}(-1)^{k}J_{2k+1}\left(\frac{2\pi E_{C}}{q}\right)\cos{\left[(2k+1)\tau\right]}\notag
    \\
    &+\frac{e^{\gamma}m}{q}\left(J_{0}\left(\frac{2\pi E_{C}}{q}\right)\right.\notag
    \\
    &\qquad\left.
    +\sum_{k=1}^{\infty}(-1)^{k}J_{2k}\left(\frac{2\pi E_{C}}{q}\right)\cos{\left[2k\tau\right]}\right).
\end{align}
In order for $\phi_{1}(\tau)$ to remain periodic instead of growing indefinitely, the first term of the RHS of Eq.~(\ref{eq:ODE for phi 1 2}) needs to vanish. 
This condition leads to the determination of the correction to the frequency as
\begin{align}
    \label{eq:omega1}
    \omega_{1}=\frac{mqe^{\gamma}}{\pi E_{C}}\cos{\left(\frac{2\pi E_{C}}{q}\right)}J_{1}\left(\frac{2\pi E_{C}}{q}\right).
\end{align}
Combining Eq.~(\ref{eq:omega1}) with Eq.~(\ref{eq:omega expand}), we obtain Eq.~(\ref{eq: plasma frequency}). 

\begin{figure}[t]
\centering
    \includegraphics[keepaspectratio, scale=0.5]{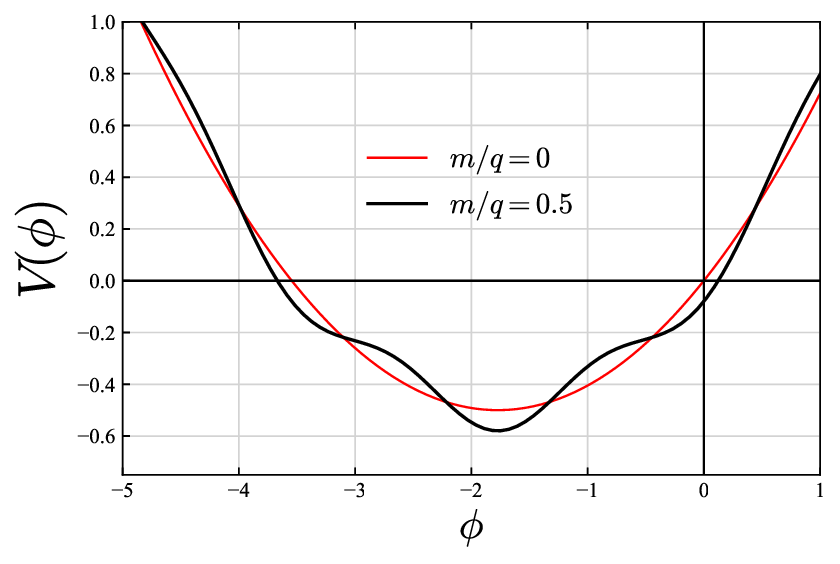}
  \caption{
  \justifying
  The potential $V(\phi)$ given in Eq.~(\ref{eq;potential}) in the case of constant $E_{C}$.  The massless and massive cases are shown by the red and black lines.
  Under this potential, $\phi$ oscillates back and forth around its minimum. For the massless case, the frequency is given by $\omega=M$. For the massive case, the potential is modified due to the third term in Eq.~(\ref{eq;potential}), and the frequency is corrected to be Eq.~(\ref{eq: plasma frequency}). 
We set $m/q=0.5$, and $E_{C}/q=1$ for this plot.
  \label{fig:potnetial}}
\end{figure}
The leading-order plasma frequency $\omega=M=q/\sqrt{\pi}$ depends only on the coupling $q$.  The first-correction \eqref{eq:omega1} displays dependence on the strength of the electric field, both through the small parameter $m/E_C$ and via the finite ratio $E_C/q$.  Notice that there are special values of $E_C/q$ where the correction $\omega_1$ vanishes entirely.  It is notable that $\omega_1$ is purely real: there is no dissipation at this order.  This fact can be attributed to the conserved quantity discussed in Sec.~\ref{sec:conserved quantity} and was already observed in our numerical study of capacitor breakdown in Sec.~\ref{sec: capacitor breakdown}.

The effect of a nonzero mass $m$ can be nicely visualized by considering a potential $V(\phi)$. First, let us write $\braket{\phi}\to \phi$ to simplify the notation. 
Then, the equation (\ref{eq:ODE for <phi>}) can be put in the following form
\begin{align}
    \label{eq:eq phi conservative}
    \partial_{t}^{2}\phi=-\frac{\partial V(\phi)}{\partial \phi},
\end{align}
where the potential $V(\phi)$ is given by
\begin{align}\label{eq;potential}
    V(\phi)=\frac{1}{2}M^{2}\phi^{2}+\frac{q}{\sqrt{\pi}}E_{C}\phi-\frac{e^{\gamma}mM}{2\pi}\cos{\sqrt{4\pi}\phi}.
\end{align}
Fig.~\ref{fig:potnetial} shows the potential $V(\phi)$ for both massless and massive cases. The overall parabolic shape of the potential comes from the first term of Eq.~(\ref{eq;potential}) and the shift of the minimum of $V(\phi)$ is due to the electric field $E_{C}$. 
The main difference between the massless and massive is the cosine term in Eq.~(\ref{eq;potential}), which can be seen as adding a slight wiggle to the parabolic line in Fig.~\ref{fig:potnetial}. 
The oscillating behavior of $\phi$ is analogous to the 1-dimensional mechanics problem. From this perspective, Eq.~(\ref{eq:eq phi conservative}) effectively describes a particle trapped in a potential (\ref{eq;potential}).   
Since any bounded motion in 1D is periodic \cite{Sideris2013}, $\phi$ oscillates back and forth in this potential and the dynamics of $\braket{\phi}$ are perfectly periodic. 

\section{Semiclassical approximation}\label{sec:QED and semiclassical}

In this section, we discuss the validity of the semiclassical approximation (e.g. \cite{Kluger.Eisenberg.ea1991oct,Kluger.Eisenberg.ea1993jun,Kluger.Mottola.ea1998nov,Pla.Newsome.ea2021may}).  We first derive the prediction of the semiclassical approximation for the problem under consideration and then compare to the full QED result.

In the semiclassical approximation, the electric field is no longer treated as an operator but just a classical field. 
This classical field creates a non-zero current $\braket{J}_{\rm ren}$ through the Schwinger mechanism.  If the current can be expressed as a functional of the classical field, i.e. $\braket{J}_{\rm ren}[E]$, then the natural Maxwell equation is
\begin{align}
    \label{eq:semiclassical approximation}
    -\partial_{t}E=\braket{J}_{\rm ren}[E]+J_{\rm ext},
\end{align}
where $J_{\rm ext}$ is the current associated with any external electric field $-\partial_t E_{\rm ext}=J_{\rm ext}$.\footnote{Previously in the paper, the external field current was denoted $E_C$ for ``classical''; here we use the label ``ext'' since all electric fields are classical in the semiclassical approximation.}   Eq.~\eqref{eq:semiclassical approximation} is the semiclassical approximation.  Since the renormalized current in general depends on the whole past history of $E$, the semi-classical equation is an integro-differential equation (or worse).

We can obtain a definite equation in the case of spatially homogeneous electric fields, where the current expectation value may be derived from the adiabatic regularization method (e.g. \cite{Ferreiro:2018qzr,Ferreiro.Navarro-Salas2018jun,Beltran-Palau:2020hdr,Pla:2022spt}). 
Following \cite{Ferreiro.Navarro-Salas2018jun}, the renormalized current density is given by 
\begin{align}
    \label{eq:<J>fermi}
    \braket{J}_{\rm{ren}}
    =&-\frac{q^{2}}{\pi}A(t)+\frac{q}{2\pi}\int^{\infty}_{-\infty}dk \left[|h^{I}_{k}|^{2}-|h^{II}_{k}|^{2}+\frac{k}{\omega}\right],
\end{align}
where  $h^{I}_{k}(t)$ and $h^{II}_{k}(t)$ are mode functions satisfying the following ODE:
\begin{align}
    \label{eq: ODE for modes}
    \frac{d}{dt}\begin{pmatrix}
        h^{I}_{k}(t)
        \\
        h^{II}_{k}(t)
    \end{pmatrix}
    =
    i\begin{pmatrix}
        k-qA(t) &-m 
        \\
        -m &-k+qA(t)
    \end{pmatrix}
    \begin{pmatrix}
        h^{I}_{k}(t)
        \\
        h^{II}_{k}(t)
    \end{pmatrix}.
\end{align}
Here, the gauge is taken to be $A^{\mu}(t)=(0,A(t))$. 
Eq.~(\ref{eq: ODE for modes}) ensures that the following normalization condition is constant in time,
\begin{align}
    \label{eq:hIhIInorm}
    |h^{I}_{k}|^{2}+|h^{II}_{k}|^{2}=1.
\end{align} 

To compare the semiclassical approximation with our QED result at linear order in $m$, we need the expression of the current Eq.~(\ref{eq:<J>fermi}) to this order.  We consider an identical setup to our full QED calculation, where the electric field is turned on at $t=t_0$.  Prior to $t=t_0$, the mode functions are those of the free theory, 
\begin{align}
    \label{eq:hIini}
    h^{I}_{k}(t<t_{0})=& \sqrt{\frac{\omega_{k,m}-k}{2\omega_{k,m}}}e^{-i\omega_{k,m} (t-t_{0})},
    \\
    \label{eq:hIIini}
    h^{II}_{k}(t<t_{0})=& -\sqrt{\frac{\omega_{k,m}+k}{2\omega_{k,m}}}e^{-i\omega_{k,m} (t-t_{0})},
\end{align}
where $\omega_{k,m}$ is given by Eq.~(\ref{eq:frequency}).  Expanding in $m$ gives 
\begin{align}
    h^{I}_{k}(t<t_0) \approx \theta(-k)e^{ik(t-t_{0})}
    +\theta(k)\frac{m}{2k}e^{-ik(t-t_{0})},
    \label{eq:hIinim} \\
    h^{II}_{k}(t<t_0) \approx \theta(k)e^{-ik(t-t_{0})}
    +\theta(-k)\frac{m}{2k}e^{ik(t-t_{0})},\label{eq:hIIinim}
\end{align}
where $\theta(k)$ is a step function. 
We expand $h^{I}_{k}(t)$ and $h^{II}_{k}(t)$ in $m$ as
\begin{align}
    \label{eq:hIexpand}
    h^{I}_{k}(t)=&x^{(0)}(t)+x^{(1)}(t)+\cdots,
    \\
    \label{eq:hIIexpand}
     h^{II}_{k}(t)=&y^{(0)}(t)+y^{(1)}(t)+\cdots,
\end{align}
where $x^{(n)}$ and $y^{(n)}$ are order $m^{n}$. 
At $t<t_{0}$, Eqs.~(\ref{eq:hIexpand}) and (\ref{eq:hIIexpand}) have to reduce to Eqs.~(\ref{eq:hIinim}) and (\ref{eq:hIIinim}); thus, we have the following conditions at $t<t_{0}$:
\begin{align}
    \label{eq:x0ini}
    x^{(0)}(t)=&\theta(-k)e^{ik(t-t_{0})},
    \\
    \label{eq:y0ini}
    y^{(0)}(t)=&-\theta(k)e^{-ik(t-t_{0})},
    \\
    \label{eq:x1ini}
    x^{(1)}(t)=&\theta(k)\frac{m}{2k}e^{-ik(t-t_{0})},
    \\
    \label{eq:y1ini}
    y^{(1)}(t)=&\theta(-k)\frac{m}{2k}e^{ik(t-t_{0})}.
\end{align}
By plugging Eqs.~(\ref{eq:hIexpand}) and (\ref{eq:hIIexpand}) into Eq.~(\ref{eq: ODE for modes}) and collecting terms order by order, we obtain the hierarchy of equations. 
From the zeroth order, we have
\begin{align}
    \label{eq:x0ODE}
    \dot{x}^{(0)}=&i(k-qA(t))x^{(0)},
    \\
    \label{eq:y0ODE}
    \dot{y}^{(0)}=&i(k-qA(t))y^{(0)}.
\end{align}
For the first order,
\begin{align}
    \label{eq:x1ODE}
    \dot{x}^{(1)}=&i(k-qA(t))x^{(1)}+i m y^{(0)},
    \\
    \label{eq:y1ODE}
    \dot{y}^{(1)}=&i(k-qA(t))y^{(1)} +i m x^{(0)}.
\end{align}
We can solve Eqs.~(\ref{eq:x0ODE}) and (\ref{eq:y0ODE}) immediately and find 
\begin{align}
    \label{eq:x0sol}
    x^{(0)}(t)=&\theta(-k)e^{i\chi(t,t_{0})}
    \\
    \label{eq:y0sol}
     y^{(0)}(t)=&-\theta(k)e^{-i\chi(t,t_{0})}
\end{align}
where $\chi(t,t_{0})$ is defined as
\begin{align}
    \label{eq:definechi}
    \chi(t,t_{0})=\int_{t_{0}}^{t}(k-qA(t'))dt'.
\end{align}
It is easy to check that Eqs.~(\ref{eq:x0sol}) and (\ref{eq:y0sol}) indeed satisfy Eqs.~(\ref{eq:x0ini}) and (\ref{eq:y0ini}) at $t<t_{0}$, respectively. 
The calculation for the first-order term is the same and the solutions to Eqs.~(\ref{eq:x1ODE}) and (\ref{eq:y1ODE}) satisfying Eqs.(\ref{eq:x1ini}) and (\ref{eq:y1ini}) are given by
\begin{align}
    \label{eq:x1sol}
    x^{(1)}(t)=&-im\theta(k)e^{i\chi(t,t_{0})}\left(\int^{t}_{t_{0}}e^{-2i\chi(s,t_{0})}ds-\frac{1}{2ik}\right),
    \\
    \label{eq:y1sol}
    y^{(1)}(t)=&im\theta(-k)e^{-i\chi(t,t_{0})}\left(\int^{t}_{t_{0}}e^{2i\chi(s,t_{0})}ds+\frac{1}{2ik}\right).
\end{align}
Expanding the integrand of \eqref{eq:<J>fermi} to order $m$ gives
\begin{align}
    |h^{I}_{k}|^{2}&-|h^{II}_{k}|^{2}+\frac{k}{\omega} \approx |x^{(0)}|^{2}-|y^{(0)}|^{2}+\frac{k}{|k|} \notag
     \\
     & \qquad \qquad
     +2\text{Re}\left(x^{(0)*}x^{(1)}-y^{(0)*}y^{(1)}\right).
\end{align}
However, the RHS actually vanishes.  The first line vanishes after using the explicit expressions \eqref{eq:x0sol} and \eqref{eq:y0sol}, while the second line vanishes because $x^{(0)}$ \eqref{eq:x0sol} has support only for positive $k$, while $x^{(1)}$ \eqref{eq:x1sol} has support only for negative $k$ (and similarly for $y$). Thus, accurate to $O(m)$, we find 
\begin{align}
    \label{eq:<J>fermiresult}
    \braket{J}_{\rm{ren}}
    =&-\frac{q^{2}}{\pi}A(t),
\end{align}
This is just the massless result, so the semiclassical approximation predicts \textit{no correction} to order $m$.  Taking a time derivative of \eqref{eq:semiclassical approximation} and using \eqref{eq:<J>fermiresult} as well as $E=-\partial_t A$ and $J_{\rm ext}=-\partial_t E_{\rm ext}$, we find
\begin{align}\label{eq:PDE for <E>}
    \left(\partial_{t}^2+\frac{q^{2}}{\pi}\right)E=\partial_t^2 E_{\rm ext}.
\end{align}
This is the semiclassical prediction, consistent to order $m$, in the case of spatially homogeneous electric fields.  By contrast, the fully quantum result \eqref{eq:PDE for <phi>}-\eqref{eq:<E>} in this case predicts\footnote{We let $E \to \widehat{E}$ and $E_C \to E_{\rm ext}$ to match the notation of this section and emphasize that the electric field is an operator in the fully quantum treatment.} 
\begin{align}
    \label{eq:PDE for <E> semiclassical}
    &\left(\partial_{t}^{2}+\frac{q^{2}}{\pi}\right)\braket{\widehat{E}}+\frac{e^{\gamma}mq^2}{\pi^{3/2}}\sin\frac{2\pi\left(\braket{\widehat{E}}-E_{\rm ext}\right)}{q}=\partial_t^2 E_{\rm ext}.
\end{align}
Comparing \eqref{eq:PDE for <E>} and \eqref{eq:PDE for <E> semiclassical}, we see that the semiclassical approximation agrees exactly with the fully quantum expectation  value at $m=0$, but misses the sine term at $O(m)$.  In particular, it fails to capture the shift in plasma frequency discussed in Sec.~\ref{sec:plasma frequency}.  The exactness of the semiclassical approximation for a massless fermion in a spatially homogeneous electric field was noticed previously in \cite{Pla.Newsome.ea2021may,Navarro-Salas.Pla2022nov}.  Here we demonstrate the precise manner in which it fails at the first order in $m$.

\section{Outlook}\label{sec:outlook}

We conclude by mentioning some open questions and interesting future directions.

Our main result is that the expectation value of the electric field obeys a partial differential equation through $O(m)$.  An obvious question is whether such a simple description continues to hold at higher order---will $\braket{E}$ still obey a PDE, or will there be non-local effects?  If the perturbation series is smooth in $m$, then one can answer this question by evaluating higher order terms in the Dyson series \eqref{eq:<phi> int picture} and \eqref{eq:U int}.  On the other hand, in the semiclassical approximation there are $m^2 \log m$ corrections that require an alternative approach \cite{paper2}, and similar difficulties may present in full QED.  If a PDE still appears at the next order, it raises the intriguing possibility that $\braket{E}$ actually obeys local dynamics in the exact theory.

Another feature of our $O(m)$ result is the lack of dissipation, in the sense that oscillations do not damp out (they can only decay due to spreading out).  Mathematically, this feature stems from the conserved energy $\mathcal{E}$ \eqref{eq: total energy} associated with the PDE.  Physically, we may say that energy is exchanged between the electric field and the matter, oscillating back and forth indefinitely.  However, the semiclassical approximation for finite fermion mass does show secular decrease in the strength of the electric field \cite{Kluger.Eisenberg.ea1991oct,Kluger.Eisenberg.ea1992jun,Kluger.Eisenberg.ea1993jun,Kluger.Mottola.ea1998nov,Pla:2020tpq,Newsome:2024spi}, which can be regarded as permanent transfer of energy from the field to the matter.  
It is not clear at present whether this electric field dissipation is an artifact of the semiclassical approach, or whether the lack of dissipation in our result is an artifact of the first-order expansion in $m$.

It is difficult to predict the answer in advance, since we lack a clear physical picture of the origin of the oscillations.  In particular, it would be desirable to have more clarity about the extent to which ``collective particle behavior'' is a useful physical picture.  Can the dynamics of 1+1D strong-field QED be profitably understood in terms of a large number of point particles interacting with a field that can create and move them, or do these results demand abandonment of this cherished idea?  Further research is required to clarify the issue.

Another interesting future direction is comparison with lattice simulations of the theory.  In particular, Ref.~\cite{Nagano:2023uaq} observed field oscillations attributed to pair production.  A larger parameter study with longer-time evolution would allow a direct comparison with the perturbative results we report.

It is also interesting to ask about the evolution of the field when the initial state is not the vacuum. In the massless case studied in \cite{Chu.Vachaspati2010apr}, the field equation is linear, and the expectation value of the field in any state always satisfies the classical Klein-Gordon equation.  By contrast, our perturbative result for the massive case applies only to evolution from the vacuum state.  A particularly interesting case is when the initial state is mixed.  By choosing a thermal state as the initial condition,  one may be able to capture familiar plasma effects, such as Landau damping, in the setting of 1+1D massive QED.

Finally, 1+1D massive QED might give some insight into astrophysical phenomena around pulsars and black holes surrounded by a strong magnetic field. Since the ultra-strong magnetic field forces charged particles to move along the direction of the magnetic field, treating the matter field as 1+1D becomes justifiable \cite{Thompson:1998ss,Gralla2019maya,Maldacena:2020skw,witten2025introductionblackholethermodynamics}. 
Moreover, the modification of the Coulomb law in the presence of a strong magnetic field \cite{Shabad:2007zu} suggests that some kind of dimensional reduction may also occur in the electromagnetic sector.  If an effective 1+1D theory indeed arises in strong magnetic fields, then the bosonization approach offers the attractive possibility of capturing complicated plasma dynamics with a simple PDE.  Indeed, the polar cap discharge for pulsars \cite{Timokhin2010nov,Timokhin.Arons2013feb,Philippov:2020jxu,Tolman:2022unu} and analogous gap formation and screening in black hole magnetospheres \cite{Levinson.Cerutti2018aug,Chen.Yuan2020jun,Crinquand:2020ppq} are astrophysical examples of electric field breakdown due to pair creation.  It would be fascinating to study these important astrophysical processes as 1+1D QED phenomena.

\section*{Acknowledgments}
It is a pleasure to acknowledge Paul Anderson for numerous helpful discussions. 
This work was supported by grants from the Simons Foundation (MP-SCMPS-00001470) and the National Science Foundation (PHY-2309191, PHY-2513082).

\appendix

\section{Interaction picture for a system with a time-varying Hamiltonian}\label{app:interactting picutre}
In this Appendix, we briefly review time-dependent perturbation theory using the interaction picture. Although this perturbation method is almost identical to the standard one \cite{Peskin:1995ev,Itzykson:1980rh,Schwartz:2014sze}, the explicit time-dependence of the Hamiltonian slightly modifies the definitions of some operators.

We consider a Schr\"odinger-picture Hamiltonian $\widehat{H}_{T}(t)$ that may have explicit time-dependence, as in Eq.~(\ref{eq:total Hamiltonian}).  
The physical state $\ket{\psi(t)}$ in the Schr\"odinger picture evolves with time, and its time evolution can be written as a linear transformation of $\ket{\psi(t)}$ by the unitary time evolution operator $\widehat{U}_{T}(t,t')$ through $\ket{\psi(t)}=\widehat{U}_{T}(t,t')\ket{\psi(t')}$. 
$\widehat{U}_{T}(t,t')$ satisfies the Schr\"odinger eqaution
\begin{align}
    \label{eq:define U(t,t')}
    i\frac{\partial}{\partial t}\widehat{U}_{T}(t,t')=&\widehat{H}_{T}(t)\widehat{U}_{T}(t,t')
\end{align}
with the condition $\widehat{U}_{T}(t,t)=\widehat{1}$. 
$\widehat{H}_{T}(t)$ can be split into the ``free" part and the interaction part: 
\begin{align}
    \widehat{H}_{T}(t)=\widehat{H}_{0}(t)+\widehat{H}_{\rm int}(t).
\end{align}
Using $\widehat{H}_{0}(t)$, the ``free'' evolution operator $\widehat{U}_{0}(t,t')$ is defined as the solution to  
\begin{align}
    \label{eq:define U0(t,t')}
    i\frac{\partial}{\partial t}\widehat{U}_{0}(t,t')=&\widehat{H}_{0}(t)\widehat{U}_{0}(t,t'),
\end{align}
with $\widehat{U}_{0}(t,t)=\widehat{1}$.  Given a Schr\"odinger-picture operator $\widehat{O}^{S}(t)$ (possibly with explicit time-dependence), the interaction-picture version $\widehat{O}^{I}(t)$ is defined
\begin{align}
    \label{eq:define interaction O}
    \widehat{O}^{I}(t)=\widehat{U}_{0}^{\dagger}(t,t_{0})\widehat{O}^{S}(t)\widehat{U}_{0}(t,t_{0}),
\end{align}
where $t_0$ is an arbitrary reference time.  The time evolution of $\widehat{O}^{I}(t)$ is then given by the Heisenberg equation that only involves the "free" part of the theory,
\begin{align}
    \label{eq:heisenberg O interaction}
    \frac{\partial}{\partial t}\widehat{O}^{I}(t)=&\frac{1}{i}\left[\widehat{O}^{I}(t),\widehat{H}_{0}^{I}(t)\right]+\left(\frac{\partial\widehat{O}(t)}{\partial t}\right)^{I},
\end{align}
where 
\begin{align}
    \left(\frac{\partial\widehat{O}(t)}{\partial t}\right)^{I}\equiv 
    \widehat{U}_{0}^{\dagger}(t,t_{0})\frac{\partial \widehat{O}^{S}(t)}{\partial t}\widehat{U}_{0}(t,t_{0}).
\end{align}
The ``interaction'' time evolution operator $\widehat{U}_{\rm int}(t,t')$ is defined by 
\begin{align}
    \label{eq:define Uint(t,t')}
    \widehat{U}_{\rm int}(t,t')=\widehat{U}_{0}^{\dagger}(t,t_{0})\widehat{U}_{T}(t,t')\widehat{U}_{0}(t',t_{0}),
\end{align}
and satisfies $\widehat{U}_{\rm int}(t,t)=\hat{1}$.  
By taking the time derivative of Eq.~(\ref{eq:define Uint(t,t')}) and using Eqs.(\ref{eq:define U0(t,t')}) and (\ref{eq:heisenberg O interaction}), the time evolution of $\widehat{U}_{\rm int}(t,t')$ is obtained as
\begin{align}
    \label{eq:schw Uint}
    i\frac{\partial}{\partial t}\widehat{U}_{\rm int}(t,t')=&\widehat{H}^{I}_{\rm int}(t)\widehat{U}_{\rm int}(t,t').
\end{align}
The formal solution to Eq.(\ref{eq:schw Uint}) can be expressed as the Dyson series,
\begin{align}
    \label{eq:Uint dyson series}
    \widehat{U}_{\rm int}(t,t')=T\exp{\left(-i\int^{t}_{t'}\widehat{H}^{I}_{\rm int}(s)ds\right)},
\end{align}
where $T$ is the time-ordering operator.

\section{Energy expectation value}\label{app:energy expectation value}
In this Appendix, we show that the conserved quantity (\ref{eq: total energy}) is equal to the expectation value of the Hamiltonian operator (\ref{eq:Hamiltonian QED boson}), after the non-dynamical part $\frac{1}{2}E_{C}^{2}$ is added back in. 
Using Eq.~(\ref{eq: NM commutes with U0}) to commute $\widehat{U}_{0}(t,,t')$ and the normal ordering $N_{M}\{\cdots\}$, the Hamiltonian operator in the interaction picture becomes 
\begin{align}
     \label{eq: total hamiltonian interaction pre}
    &\widehat{H}^{I} (t)
    =\int dxN_{M}\left\{\frac{1}{2}(\widehat{\Pi}^{I})^{2}+\frac{1}{2}\left(\partial_{x}\widehat{\phi}^{I}\right)^{2}
    +\frac{1}{2}M^{2}(\widehat{\phi}^{I})^{2}\right.\notag
    \\
    &\left.
    +\frac{q}{\sqrt{\pi}}E_{C}\widehat{\phi}^{I}+\frac{1}{2}E_{C}^{2}
    -\frac{e^{\gamma}mM}{2\pi}\cos{\sqrt{4\pi}\widehat{\phi}^{I}}
    \right\},
\end{align}
with $\frac{1}{2}E_{C}^{2}$ added back in.  
After using Eqs.~(\ref{eq:op sol phi interaction}) and (\ref{eq:op sol pi interaction}) for the first three terms of Eq.~(\ref{eq: total hamiltonian interaction pre}),  one may split Eq.~(\ref{eq: total hamiltonian interaction pre}) into the following:
\begin{align}
    \widehat{H}^{I} (t)\
     \label{eq: total hamiltonian interaction}
    &= \widehat{H}^{I}_{1}+\widehat{H}^{I}_{2}+\widehat{H}^{I}_{3}+\widehat{H}^{I}_{4}
\end{align}
where each piece is defined by 
\begin{align}
    \label{eq:H1}
    \widehat{H}^{I}_{1}&=\int \frac{dk}{2\pi} \omega_{k,M}
    \widehat{a}^{\dagger}_{k,M}\widehat{a}_{k,M},
    \\
    \label{eq:H2}
    \widehat{H}^{I}_{2}&=\int dx\left[\partial_{t}\widehat{\phi}^{I}\partial_{t}\phi_{\rm cl}+\partial_{x}\widehat{\phi}^{I}\partial_{x}\phi_{\rm cl}
    +M^{2}\widehat{\phi}^{I}\phi_{\rm cl}\right]\notag
    \\
     &- \int dx\left[\frac{1}{2}(\partial_{t}\phi_{\rm cl})^{2}+\frac{1}{2}\left(\partial_{x}\phi_{\rm cl}\right)^{2}
    +\frac{1}{2}M^{2}\phi_{\rm cl}^{2}\right],
    \\
    \label{eq:H3}
     \widehat{H}^{I}_{3}&= \int dx \left(\frac{q}{\sqrt{\pi}}\widehat{\phi}^{I}E_{C}+\frac{1}{2}E_{C}^{2}\right),
     \\
     \label{eq:H4}
     \widehat{H}^{I}_{4}&=\int 
    N_{M}\left\{-
    \frac{e^{\gamma}mM}{2\pi}\cos{\sqrt{4\pi}\widehat{\phi}^{I}}\right\}dx.
\end{align}

Now, we evaluate the expectation value of Eqs.~(\ref{eq:H1})--(\ref{eq:H4}) piece by piece. 
To see the expectation vale of Eq.~(\ref{eq:H1}), consider the expectation value of $\widehat{a}^{\dagger}_{k,M}\widehat{a}_{k,M}$.  
Using Eqs.~(\ref{eq: formaula for |t,Omega>}), (\ref{eq: NM commutes with U0}), (\ref{eq:cos phi commu interaction}), and (\ref{eq:op to cl relatio}) and keeping to first order, we find
\begin{align}
    &\braket{\widehat{a}^{\dagger}_{k,M}\widehat{a}_{k,M}}\notag
    \\
    &=
    \bra{0,M}\widehat{U}_{\rm int}^{\dagger}(t,-\infty)\widehat{a}^{\dagger}_{k,M}\widehat{a}_{k,M}\widehat{U}_{\rm int}(t,-\infty)\ket{0,M}\notag
    \\
    &\approx
     \bra{0,M}\widehat{a}^{\dagger}_{k,M}\widehat{a}_{k,M}\ket{0,M}
    \notag
    \\
    &+\int_{-\infty}^{t}dt'\int\!dy
    \left(-i\frac{e^{\gamma}mMf(t')}{2\pi\sqrt{\pi}}\right)\notag
    \\
   & \times
    \bra{0,M}\left[N_{M}\left\{\cos{\sqrt{4\pi}\widehat{\phi}^{I}(t',y)}\right\},\widehat{a}^{\dagger}_{k,M}\widehat{a}_{k,M}\right]\ket{0,M}\notag
    \\
    &=0,
\end{align}
since both 0th and 1st order terms vanish due to $\widehat{a}_{k,M}\ket{0,M}=0$ (equivalently $\bra{0,M}\widehat{a}^{\dagger}_{k,M}=0$). 
This leads to 
\begin{equation}
    \label{eq: H1 expectation value}
    \braket{H_{1}}=0.
\end{equation}

For the computation of $\braket{H_{2}}$, notice that operator $\widehat{U}_{\rm int}(t,-\infty)$ does not depend on the spatial coordinate. Thus, using Eq.~(\ref{eq:<phi> int picture}), we find
\begin{align}
    \partial_{x}\braket{\phi}&=\braket{\partial_{x}\widehat{\phi}^{I}}. 
\end{align}
For the time derivative of $\braket{\phi}$, one may use Eqs.~(\ref{eq:<phi> int picture}) and (\ref{eq:schw Uint}), then
\begin{align}
    \partial_{t}\braket{\phi}
    &=\braket{\partial_{t}\widehat{\phi}^{I}}+i\Braket{\left[\widehat{H}^{I}_{\rm int}(t),\widehat{\phi}^{I}(t,x)\right]}\notag
    \\
    &=\braket{\partial_{t}\widehat{\phi}^{I}}, 
\end{align}
where the commutator of $\widehat{H}^{I}_{\rm int}$ and $\phi^{I}(t,x)$ vanishes due to  the equal time commutation relation (\ref{eq:equal time phi phi commu interaction}).  
Using these results and the fact that $\braket{\phi}=\phi_{\rm cl}+O(m)$, the expectation value of Eq.~(\ref{eq:H1}) to first order in $m$ becomes
\begin{align}
    \label{eq: H2 expectation value}
    \braket{\widehat{H}_{2}}
    &=
    \int dx \left(\frac{1}{2}(\partial_{t}\braket{\phi})^{2}+\frac{1}{2}(\partial_{x}\braket{\phi})^{2}+\frac{1}{2}M^{2}\braket{\phi}^{2}\right).
\end{align}
Furthermore, the expectation value of Eq.~(\ref{eq:H3}) is simply
\begin{align}
    \label{eq: H3 expectation value}
    \braket{\widehat{H}_{3}}&=\int dx \left(\frac{q}{\sqrt{\pi}}\braket{\phi}E_{C}+\frac{1}{2}E_{C}^{2}\right),
\end{align}
since $\widehat{H}^{I}_{3}$ is only linear in $\widehat{\phi}^{I}$. 
For Eq.~(\ref{eq:H4}), notice that this term is already at order $m$, thus, we obtain
\begin{align}
   \label{eq: H4 expectation value}
     \braket{\widehat{H}_{4}}
    &=\int \left(-
    \frac{e^{\gamma}mM}{2\pi}\cos{\sqrt{4\pi}{\braket{\phi}}}\right)dx
\end{align} 

Combining Eqs.~(\ref{eq: H1 expectation value}) and (\ref{eq: H2 expectation value})--(\ref{eq: H4 expectation value}) and comparing the result to the energy $\mathcal{E}$ defined in Eq.~(\ref{eq: total energy}), one may conclude $\braket{\widehat{H}}=\mathcal{E}$
to first order in $m$, as presented in Eq.~(\ref{eq:<H>=E}) in the main text. 
In other words, the conserved quantity derived in Sec.~\ref{sec:conserved quantity} is genuinely the energy expectation value of the system. 

Even though the total energy $\mathcal{E}$ can be identified as the total energy expectation, the particle part (\ref{eq: particle energy}) and field part (\ref{eq: field energy}) are not the expectation value of their quantum counterparts. Similar to Eq.~(\ref{eq: total energy}) was split into Eqs.~(\ref{eq: particle energy}) and (\ref{eq: field energy}), one may split 
\begin{align}
    \widehat{H}=\widehat{H}_{\rm particle}+\widehat{H}_{\rm field},
\end{align}
where 
\begin{align}
    \label{eq: particle Hamiltonian}
    &\widehat{H}_{\text{particle}}\notag
    \\
    &=
    \int N_{M}\left\{\frac{1}{2}\widehat{\Pi}^{2}+\frac{1}{2}\left(\partial_{x}\widehat{\phi}\right)^{2}-
    \frac{e^{\gamma}mM}{2\pi}\cos{\sqrt{4\pi}\widehat{\phi}}\right\}dx,
\end{align}
and
\begin{align}
    \label{eq: field Hamiltonian}
    \widehat{H}_{\text{field}}
    &= \int \frac{1}{2}\widehat{E}^2dx = 
     \int\frac{1}{2}N_{M}\left\{\left(E_{C}+\frac{q}{\sqrt{\pi}}\widehat{\phi}\right)^{2}\right\}dx.
\end{align}
However, the direct calculation to order $m$ using Eqs.~(\ref{eq: formaula for |t,Omega>}),(\ref{eq: NM commutes with U0}),(\ref{eq:cos phi commu interaction}),(\ref{eq:op to cl relatio}),(\ref{eq:expectation value phi}) shows 
\begin{align}
    \label{eq:<phi^2> calc} 
    &\Braket{N_{M}\{\widehat{\phi}^{2}\}}\notag
    \\
    &=
    \bra{0,M}\widehat{U}_{\rm int}^{\dagger}(t,-\infty)N_{M}\{(\widehat{\phi}^{I})^{2}\}\widehat{U}_{\rm int}(t,-\infty)\ket{0,M},\notag
    \\
    &=\phi_{\rm cl}(t,x)^{2}\notag
    \\
    &+\int_{-\infty}^{t}dt'\int\!dy
    \left(-i\frac{e^{\gamma}mMf(t')}{2\pi\sqrt{\pi}}\right)2\phi_{\rm cl}(t,x)\notag
    \\
   & \times
    \bra{0,M}\left[N_{M}\left\{\cos{\sqrt{4\pi}\widehat{\phi}^{I}(t',y)}\right\},\widehat{\phi}^{I}_{\rm ff}(t,x)\right]\ket{0,M}+\Delta\notag
    \\
    &=\braket{\phi}^{2}+\Delta
\end{align}
where $\phi^{I}_{\rm ff}(t,x)$ is defined by 
\begin{align}
    \phi^{I}_{\rm ff}(t,x)=\phi^{I}(t,x)-\phi_{\rm cl}. 
\end{align}
and $\Delta$ is 
\begin{align}
    \label{eq:define Delta}
     &\Delta =\int_{-\infty}^{t}dt'\int\!dy
    \left(-i\frac{e^{\gamma}mMf(t')}{2\pi\sqrt{\pi}}\right)\notag
    \\
   & \times
    \bra{0,M}\left[N_{M}\left\{\cos{\sqrt{4\pi}\widehat{\phi}^{I}(t',y)}\right\},N_{M}\{\widehat{\phi}^{I}_{\rm ff}(t,x)^{2}\}\}\right]\ket{0,M},
\end{align}
However, $\Delta$ is in general nonzero, since there is no obvious annihilation of $\bra{0,M}$ or $\ket{0,M}$ by $\widehat{a}^{\dagger}_{k,M}$ and $\widehat{a}_{k,M}$ in Eq.~(\ref{eq:define Delta}).  
In other words, Eq.~(\ref{eq:<phi^2> calc}) leads to 
\begin{align}
    \label{eq:<phi^2>neq<phi>^2}
    \Braket{N_{M}\{\widehat{\phi}^{2}\}}\neq\braket{\phi}^{2}. 
\end{align}
An immediate consequence of this result is 
\begin{align}
    \label{eq: H field neq E field}
     \braket{\widehat{H}_{\rm field}}&\neq\mathcal{E}_{\rm field}, 
\end{align}
which can be quickly verified by taking the expectation value of Eq.~(\ref{eq: field Hamiltonian}) and comparing it to Eq.~(\ref{eq: field energy}) using Eq.~(\ref{eq:<phi^2>neq<phi>^2}). 
Since $\braket{\widehat{H}}=\mathcal{E}$, Eq.~(\ref{eq: H field neq E field}) also implies,
\begin{align}
    \label{eq: H particle neq E particle}
    \braket{\widehat{H}_{\rm particle}}&\neq\mathcal{E}_{\rm particle}.
\end{align}

\bibliography{ScalarField}
\bibliographystyle{utphys}

\end{document}